\documentclass[prb,showpacs,floatfix,twocolumn]{revtex4}

\usepackage{graphicx}

\newcommand{\bto}{Bi$_2$Ti$_2$O$_6$O$^\prime$}
\newcommand{\obi}{O$^\prime$Bi$_4$}
\newcommand{\obo}{O$^\prime$--Bi--O$^\prime$}

\begin{document}

\title{Atomic displacements in the ``charge-ice'' pyrochlore
Bi$_2$Ti$_2$O$_6$O$^\prime$ studied by neutron total scattering}

\author{Daniel P. Shoemaker} \email{dshoe@mrl.ucsb.edu}
\author{Ram Seshadri}
\affiliation{Materials Department and Materials Research Laboratory,
University of California, Santa Barbara, CA, 93106, USA}
\author{Andrew L. Hector} 
\affiliation{School of Chemistry, University of Southampton, Highfield, 
Southampton SO17 1BJ, UK} 
\author{Anna Llobet}
\author{Thomas Proffen}
\affiliation{Los Alamos National Laboratory, Lujan Neutron Scattering Center, 
MS H805, Los Alamos, New Mexico 87545, USA}
\author{Craig J. Fennie}
\affiliation{School of Applied and Engineering Physics, Cornell University,
Ithaca, New York 14853, USA}

\date{\today}


\begin{abstract}

The oxide pyrochlore Bi$_2$Ti$_2$O$_6$O$^\prime$ is known to be associated with large 
displacements of Bi and O$^\prime$ atoms from their ideal crystallographic 
positions. Neutron total scattering, analyzed in both reciprocal and real 
space, is employed here to understand the nature of these displacements. 
Rietveld analysis and maximum entropy methods are used to produce an average 
picture of the structural non-ideality. Local structure is modeled 
\textit{via} large-box reverse Monte Carlo simulations constrained 
simultaneously by the Bragg profile and real-space pair distribution function. 
Direct visualization and statistical analyses of these models show the precise 
nature of the static Bi and O$^\prime$ displacements. Correlations between 
neighboring Bi displacements are analyzed using coordinates from the 
large-box simulations. The framework of continuous symmetry measures has been
applied to distributions of O$^\prime$Bi$_4$ tetrahedra to examine deviations
from ideality. Bi displacements from ideal positions appear 
correlated over local length scales. The results are 
consistent with the idea that these nonmagnetic lone-pair containing pyrochlore 
compounds can be regarded as highly \textit{structurally} frustrated systems. 

\end{abstract}

\maketitle 

\section{Introduction} 

Magnetic oxides with the $A_2B_2$O$_7$ pyrochlore structure have been the 
subject of intense study. The lattice of corner-connected tetrahedra of 
$A$ atoms hinders cooperative magnetic ordering, and when the $A$ atom spins
are Ising, an ice-like ground state is produced, for example in 
Dy$_2$Ti$_2$O$_7$ and Ho$_2$Ti$_2$O$_7$.\cite{ramirez_strongly_1994,ramirez_zero-point_1999,morris_dirac_2009,kadowaki_observation_2009} 
Other recent developments in oxide pyrochlores include superconductivity in
osmium compounds,\cite{yonezawa_new_2004} the formation of polar metallic 
states with unusual phonon modes,\cite{kendziora_goldstone_2005}, and the 
suggestion of chiral magnetic ground states.\cite{onoda_spin_2003} 
The analogy between Ising spins and vector displacements of cations within 
their coordination polyhedra has led to the 
suggestion that polar ordering may be similarly frustrated on the 
pyrochlore lattice.\cite{seshadri_lone_2006} Thus in pyrochlore \bto, it is 
known that the Bi$^{3+}$ atoms, usually predisposed to off-centering within 
their coordination polyhedra, display incoherent displacements permitting
the average structure to remain cubic.\cite{hector_synthesis_2004} This is
in sharp contrast to the Bi$^{3+}$-containing perovskites BiMnO$_3$ and BiFeO$_3$
where the lone-pair active A-site produces polar, non-cubic ground
states.\cite{hill_density_2002} Some signatures of these incoherent displacements are 
seen in measurements of heat capacity of \bto\/ and related compounds at 
low temperatures.\cite{melot_large_2009}

Significant advances in describing frustrated, ice-like behavior in magnetic 
pyrochlores have been made when atomistic models are utilized to describe the 
interactions between individual $A$ sites. For example, atomistic simulation 
of magnetic spin ordering in Dy$_2$Ti$_2$O$_7$ leads to a picture of localized, 
uncompensated spins connected by strings of ordered
spins.\cite{kadowaki_observation_2009,morris_dirac_2009}
This picture is local: spin behavior is driven by connectivity, geometry,
and pairwise exchange interactions. However, it also agrees with bulk 
thermodynamic measurements: the heat capacity of the ensemble average agrees 
with experimental observations.  We show here that experimental modeling 
over multiple length scales, both atomistic and averaged, affords a view of 
the pyrochlore \bto\, where Bi \textit{displacements}, rather than spins, form a 
frustrated network on the $A$ sublattice.

\begin{figure}
\centering\includegraphics[width=7cm]{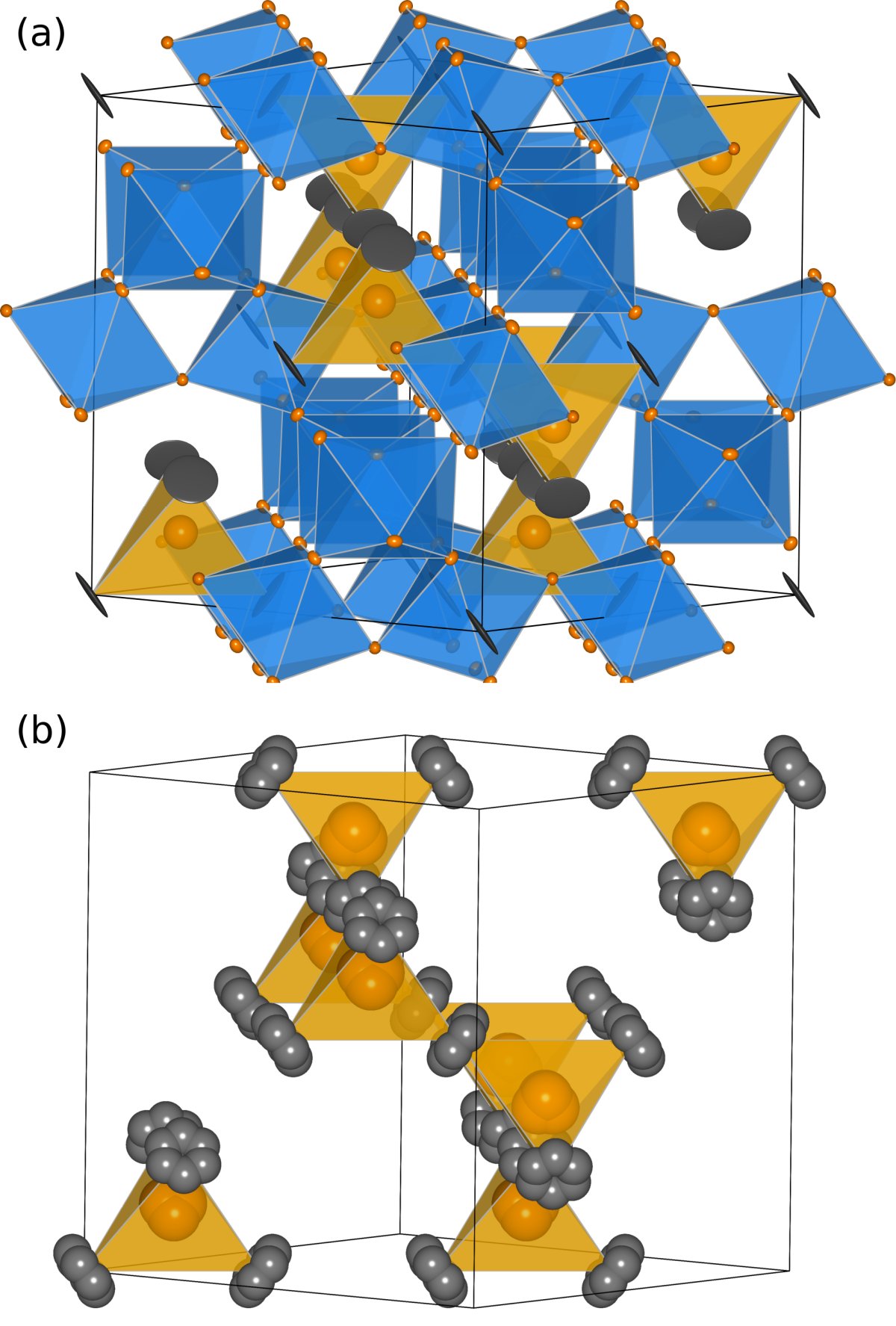} \\
\caption{(Color online) The \bto\/ crystal structure is shown in (a), with 
50\% thermal ellipsoids representing the atomic displacement parameters 
from Rietveld refinement at 14\,K. The two sublattices are
corner-sharing TiO$_6$ octahedra (blue) and corner-sharing \obi\/ tetrahedra 
(orange). Bi and O$^\prime$ are on ideal positions. Bi cations (black) appear 
as discs due to their displacement normal to the \obo\/ bond. This disorder 
can be modeled using six-fold-split Bi and four-fold-split O$^\prime$ 
as shown in (b) (only the \obi\/ is sublattice shown).}
\label{fig:unitcell}
\end{figure}

\bto\ [structure in Fig.\,\ref{fig:unitcell}(a)] is written thus to emphasize 
the two sublattices: one of corner-sharing TiO$_6$ octahedra, and the other 
of corner-sharing \obi\, tetrahedra. While the TiO$_6$ sublattice is rigid in 
models of the average structure, Bi atoms are suggested to displace 0.4\,\AA\/ 
normal to the linear \obo\,bond in an uncorrelated 
manner.\cite{hector_synthesis_2004,seshadri_lone_2006}
First-principles calculations on \bto\/ predict Bi displacements but
these are perforce associated with non-cubic 
symmetries.\cite{fennie_lattice_2007} Diffuse intensity in electron diffraction 
patterns of related compounds including 
Bi$_2$Ru$_2$O$_7$, Bi$_2$InNbO$_7$, and Bi$_2$ScNbO$_7$ may indicate
short-range correlations in the Bi displacements.\cite{withers_local_2004,goodwin_real-space_2007,liu_displacive_2009}
If Bi displacements cooperatively order with each other, they must do so only 
over short ranges. Crystallographic analysis based on Bragg scattering leaves a 
void in the ability to probe such short-range order, as
analyses are predicated on the existence of long-range order.
Consequently, studies of displacive disorder on the $A$ site \textit{via} Rietveld 
refinement or Fourier maps can produce a model of the \textit{average} electron 
or nuclear distributions, but each $A$ site has an identical cloud of 
intensity.\cite{tabira_annular_2001}

We investigate models where the 
correlated motion of atoms on the $A$ sites reproduces the atomistic, pairwise 
distances between individual atoms. This description is provided by an 
appropriate Fourier transform of the total scattering function $S(Q)$ to 
provide a normalized pair distribution function (PDF).\cite{egami_underneath_2003,proffen_analysis_2000}
In this study, the PDF and the Bragg profile are used as experimental
constraints in a large-box (11,000 atom) model of \bto\/ to obtain, using
reverse Monte-Carlo (RMC) analysis, a consistent picture of the 
the coordination tendencies of all atoms. Many of these models are combined
and used as 
a set of atomic positions for further analysis. RMC compares the
experimental and computed (based on atom positions in the simulation box) 
$D(r)$ and $S(Q)$ while randomly relaxing atomic positions. The method is
similar to Metropolis Monte Carlo, except that the fit to data $\chi^2$,
instead of a potential energy function, is minimized.\cite{
mcgreevy_reverse_1988,mcgreevy_reverse_2001,tucker_application_2001}

There are many approaches to describing the behavior and correlations of atomic
positions as obtained from large-box models of structure. Some examples 
include the use of quadrupolar moments of octahedra to describe 
LaMnO$_3$,\cite{sartbaeva_quadrupolar_2007} of contour plots of bond angles
in cristobalite,\cite{tucker_dynamic_2001} and the use of bond valence 
analysis to obtain valence states from a statistical analysis of metal-oxygen 
positions.\cite{adams_bond_2005,shoemaker_unraveling_2009}
Here we analyze the local geometry using simple metrics, then present the
continuous symmetry measures 
(CSM)\cite{zabrodsky_continuous_1992,pinsky_continuous_1998,keinan_studies_2001}
of polyhedra from RMC simulations. The CSM model provides a 
quantitative measure of a polyhedron's distortion, in the form of a 
``distance'' from ideality. A key advantage of CSM is its ability to compare 
shapes in different compounds.\cite{keinan_studies_2001} The CSM has been used 
to correlate deviations from ideal tetrahedra in silicates as a function of 
applied pressure,\cite{yogev-einot_pressure_2004} and to analyze second-order
Jahn-Teller systems across a variety of crystal 
structures.\cite{ok_distortions_2006}
In these cases, the CSM was considered for the average crystallographic 
stucture, \textit{e.g.} one where polyhedra possess a single CSM value. 
Here, we extend CSM to large-box modeling by calculating it for every 
\obi\ tetrahedron in the RMC supercell, obtaining distributions, rather than 
single values.

The key finding to emerge from this study is that displacements from ideal 
atomic positions in \bto\ and in particular, the nature of the \obi\ 
tetrahedra indicate a tendency for Bi to lie in a disordered ring around
the ideal position, with some preference for near-neighbor Bi-Bi ordering.
This reaffirms the case that, even when probed microscopically,
\bto\/ is ice-like 
in its disorder. We emphasize that in drawing the analogy with ice, we do not
suggest the existence of ice-rules of the Bernal-Fowler\cite{bernal_theory_1933} type 
in these systems.

\section{Methods}

Synthesis and a detailed average structural analysis of the sample used 
in this study (including verification of purity) has been reported by 
Hector and Wiggin.\cite{hector_synthesis_2004} Briefly, a basic solution of 
titanium metal with hydrogen peroxide and ammonia was added to an acidic 
solution of bismuth nitrate pentahydrate and nitric acid. The resulting
precipitate was filtered, washed with a dilute ammonia solution, dried at 
50$^\circ$C, and calcined in air for 16\,h at 470$^\circ$C. Time-of-flight 
(TOF) neutron powder diffraction on samples held in vanadium cans was collected
at the NPDF instrument at Los Alamos National Laboratory at 298\,K and 14\,K. 
NPDF is designed to collect high-resolution, high-momentum-transfer data 
suitable for production of the PDF, as well as traditional Rietveld 
refinement. We performed Rietveld refinement using 
GSAS.\cite{larson_general_2000}. Extraction
of the PDF with PDFGetN\cite{peterson_pdfgetn_2000} used $Q_{max}$ =
35\,\AA$^{-1}$, and least-squares refinements of the PDF were performed with
PDFgui.\cite{farrow_pdffit2_2007}

First-principles density functional methods were used to identify a
possible ordered ground state. The local stability of $Fd\bar{3}m$ Bi$_2$Ti$_2$O$_7$
was investigated using projector augmented
wave potentials within the local density approximation
as implemented in the \textsc{vasp} program\cite{kresse_ab_1993,kresse_efficient_1996,
blochl_projector_1994,kresse_ultrasoft_1999} and described previously.\cite{fennie_lattice_2007}
We found three zone-center lattice instabilities--one ferroelectric and two
antiferroelectric-like modes--which were used to guide a systematic search
for low-symmetry phases. We performed a series of structural relaxations within each of the 
highest-symmetry isotropy subgroups\cite{bilbao,isotropy} consistent with the 
freezing-in of one or more of these lattice instabilities.  We found that the lowest 
energy structure is ferroelectric (with a substantial polarization of
 $\mathbf{P} \approx$ 20\,$\mu$C\,cm$^{-2}$), in the polar monoclinic space group $Cm$,
consistent with the freezing-in of all three modes. 

Maximum entropy method (MEM) calculations of 
the nuclear scattering density were performed using 
PRIMA.\cite{izumi_recent_2002} RMC simulations were performed using 
RMCProfile\cite{tucker_rmcprofile_2007} version 6 on a 
$5 \times 5 \times 5$ supercell (cubic, $\approx 52$ \AA\/ per side) with 11000 
atoms. These simulations were constrained by the PDF (in the form of 
$D(r)$)\cite{keen_comparison_2001} up to $r$ = 12\,\AA\/ and by the Bragg 
profile of NPDF bank 1, which contains the lowest-$Q$ Bragg reflections. 
Crystal structures are visualized using VESTA\cite{momma_vesta_2008}
and AtomEye.\,\cite{li_atomeye_2003} Quantitative analyses shown in 
this work are taken from averaging many simulations in order to obtain an 
unbiased interpretation of the fit to data. Hard-sphere cutoffs were employed 
in RMC simulations to ensure that atoms did not approach closer than the
specified nearest-neighbor distances, but no bunching was observed at these
cutoffs. CSM for \obi\/ tetrahedra were calculated using a distance measure
program provided by M. Pinsky and D. Avnir.

\section{Results and Discussion}

\subsection{Average structure from Rietveld refinement}

\begin{figure}
\centering\includegraphics[width=8cm]{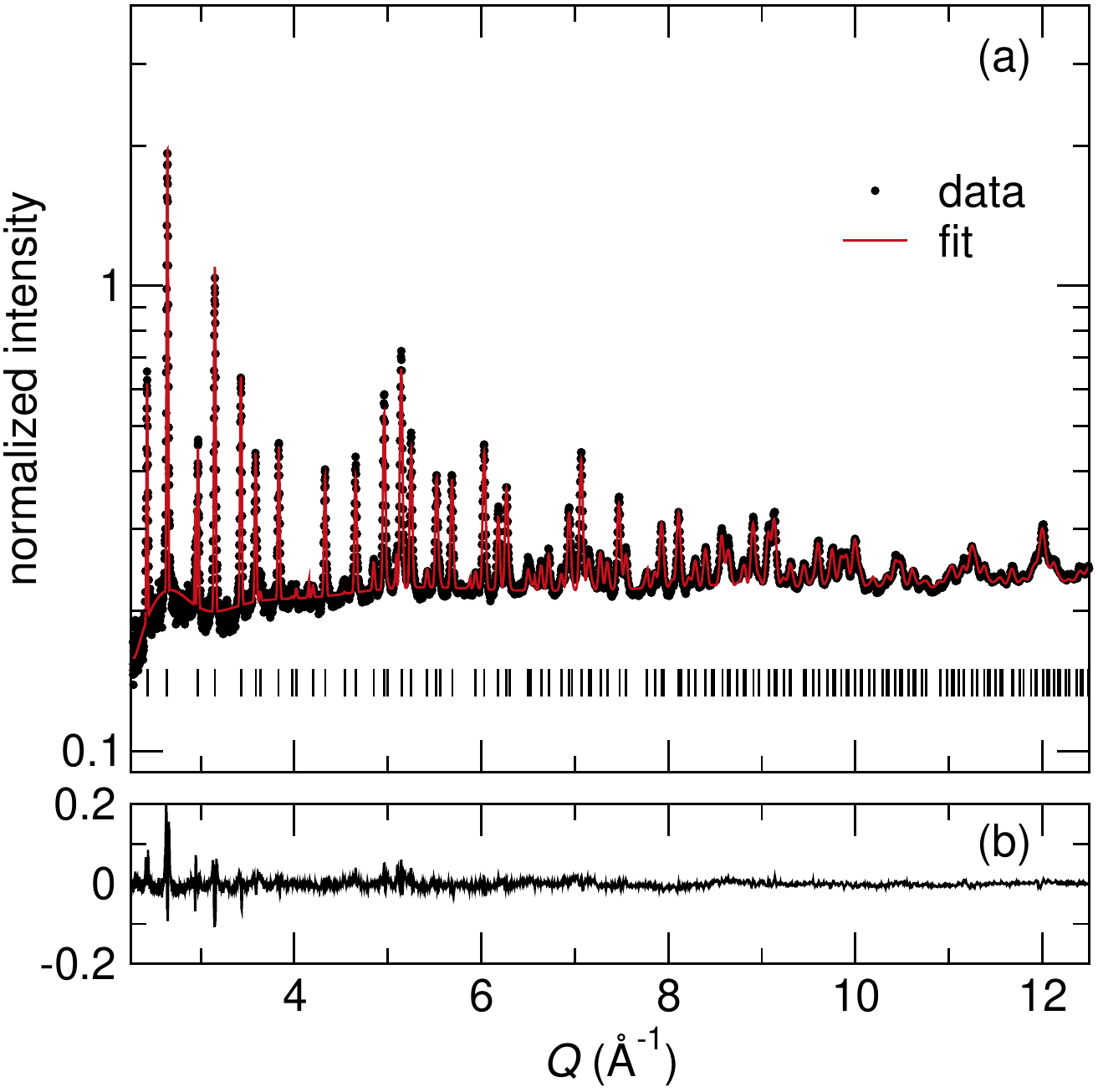} \\
\caption{(Color online) Neutron TOF Rietveld refinement of \bto\,at 14\,K (a) 
using the model of Hector and Wiggin \cite{hector_synthesis_2004} gives a
fit with $R_{wp} = 2.9\%$. Small features in the difference curve (b) at 
low $Q$ are caused by shoulders of diffuse intensity around Bragg peaks. This 
component of the data can be interpreted using total scattering analysis.}
\label{fig:gsas}
\end{figure}

The profile resulting from TOF neutron Rietveld refinement is displayed (for
a single bank of data) in Fig.\,\ref{fig:gsas}. The goodness of fit parameter 
$R_{wp}$ = 2.9\%. The sample is the same as used in the Rietveld study of
Hector and Wiggin,\cite{hector_synthesis_2004} and the fit converges to
similar structural parameters (within error), so these will not 
be reproduced here. 

In \bto, preliminary Rietveld refinement using an ideal pyrochlore model can 
obtain a good fit to data, but large atomic displacement parameters (ADPs) 
indicate that significant atomic disorder is present, characterized by an 
elliptical spread around the ideal atomic positions. The ideal Bi position on 
the $16c$ (0,0,0) site can be fit using a large $U_{23}$ component.
These appear as large, flat discs in Fig.\,\ref{fig:unitcell}(a) with a radius 
of about 0.4\,\AA. The discs envelop the Bi displacive disorder which we 
seek to accurately describe. The O$^\prime$ atoms at the $8a$ 
($\frac{1}{8}, \frac{1}{8}, \frac{1}{8}$) position have large, isotropic 
displacements as well, corresponding to $U_{iso} \approx 0.4$\/\AA$^2$. 
The TiO$_6$ sublattice, on the other hand, is described by small $U_{iso}$ 
values and does not display any signs of displacive disorder.
Here we investigate the precise nature of atomic displacements in the
\obi\,sublattice.

Improved refinement of the average pyrochlore structure has been achieved
by using a split-atom model for the $A$ sites, such as in studies of
\bto,\cite{radosavljevic_synthesis_1998,hector_synthesis_2004}
Bi$_2$Sn$_2$O$_7$,\cite{evans_alpha_2003} or La$_2$Zr$_2$O$_7$.\cite{tabira_annular_2001}
Hector and Wiggin modeled Bi using a six-fold ring in the $96g$
position,\cite{hector_synthesis_2004} but acknowledged that their refinement
does not clearly show a preference for $96g$ versus $96h$ (rotated 30$^\circ$
to each other). The $96g$ split-atom configuration is shown in 
Fig.\,\ref{fig:unitcell}(b). Comparison to an ideal-position model in 
Fig.\,\ref{fig:unitcell}(a) shows that the Bi split-atom sites lie inside the
anisotropic discs. Most O$^\prime$ intensity is still centered on the 
$8a$ site, but some occupancy is shifted away in 4 directions to form a 
tetrapod, modeled by partial occupancy of the $32e$ sites. From Rietveld 
refinement, we find that the Bi displacement parameters at 14\,K and 300\,K 
are similar in orientation and magnitude. This suggests that Bi 
displacements are frozen at room (and higher) temperatures, and what is being 
monitored in the scattering is not a snapshot of dynamic motion, but rather
a description of static positions. The suggestion of frozen displacements at 
room temperature is consistent with measurements 
of the temperature- and field-dependence of the dielectric constant
in \bto\ thin films.\cite{cagnon_microstructure_2007} 

Kinks in the difference profiles of the Rietveld refinement in 
Fig.\,\ref{fig:gsas}(b) can be attributed to shoulders of diffuse intensity 
around low-$Q$ peaks. The diffuse scattering cannot be interpreted here 
because the Rietveld technique only models Bragg intensity. Crucial 
approximations are made to model a structure using only Bragg peaks: we must 
average any atomic correlations or distortions that do not
possess long-range ordering. The Bi distortions are incoherent,
and their description will require an examination of diffuse scattering. 
Total scattering analysis of the real-space PDF, discussed here subsequently, 
provides a real-space tool to model both Bragg and diffuse scattering 
simultaneously.

\subsection{Maximum entropy method}

\begin{figure}
\centering\includegraphics[width=8cm]{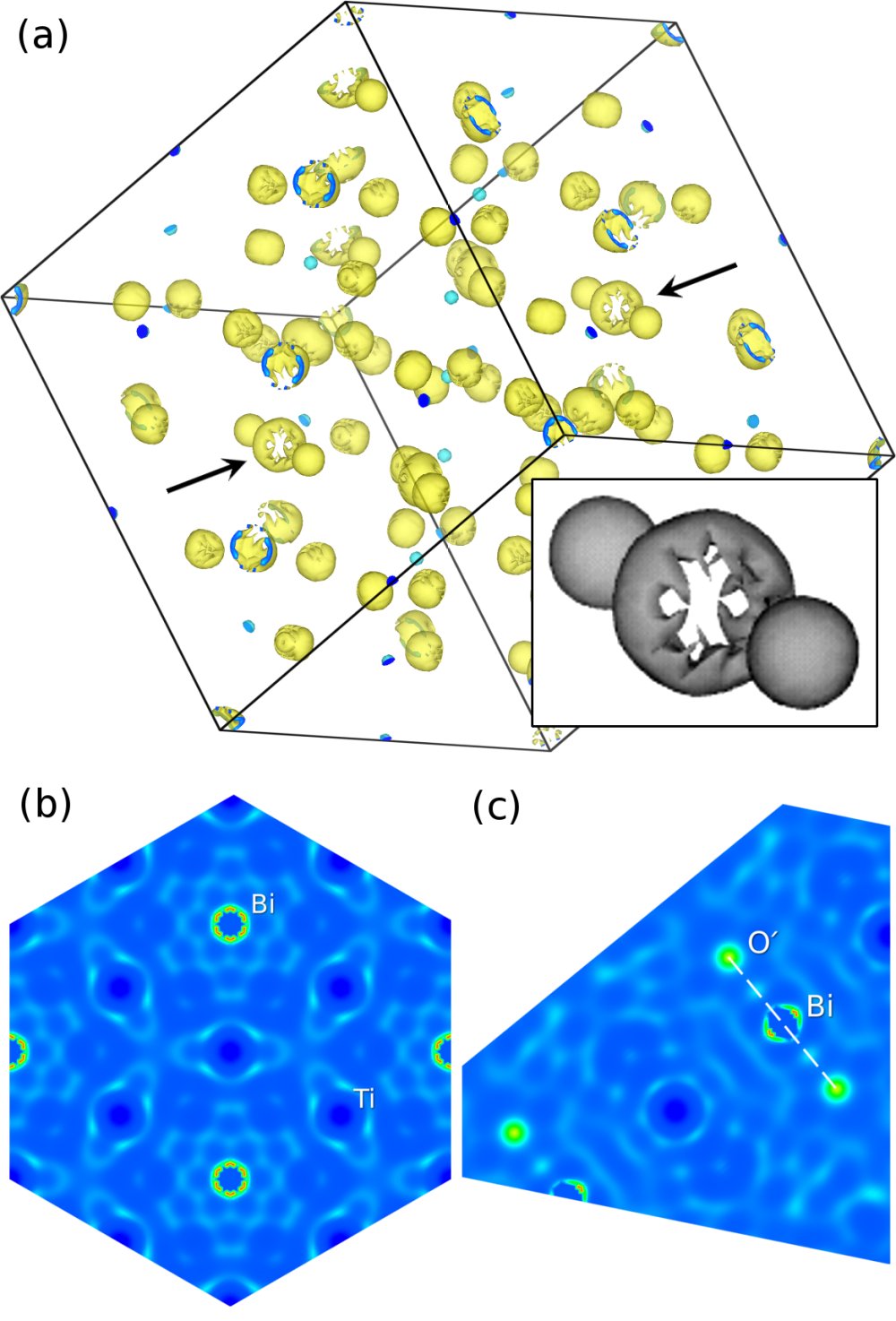}  \\
\caption{(Color online) The MEM nuclear density map of the \bto\,unit cell at 14 K is plotted
in (a). Arrows point to Bi positions where the rings are rotated with the 
\obo\/ direction almost parallel to the viewing direction. The circular Bi 
isosurface is magnified in the inset. The 2-D density slice in (b) is the
$(11\overline{1})$ plane, normal to \obo\/ bonds, and (c) is the 
$(12\overline{1})$ plane with the \obo\/ bond dashed. The imposed sixfold symmetry of the Bi ring is 
evident in the $(11\overline{1})$ slice.}
\label{fig:mem}
\end{figure}

The maximum entropy method (MEM) is used here to produce a map of neutron scattering 
density in the unit cell. The method, 
proposed originally by Jaynes,\cite{jaynes_information1_1957,jaynes_information2_1957} 
uses some testable information (in this case, observed Bragg peak 
intensities in TOF neutron diffraction), and is based on the most probable distribution 
(nuclear scattering density) being the one with the largest information entropy. 
This method is described by Sakata\cite{sakata_accurate_1990,sakata_electron_1990} 
and in the context of the MEM software program MEED,\cite{kumazawa_meed_1993}
a precursor to the software code PRIMA used here. The MEM input consists of
the observed Bragg structure factor $F_{obs}$ (from a Le Bail fit, with phase
information from the ideal structure) for a list of $hkl$ reflections, the
unit cell dimensions,  space group, and the sum of all scattering lengths
in the cell. The MEM as employed here is largely model-free in the sense
that atomic positions are  not specified during the calculation, but phase
information for $F_{obs}$ is biased by the ideal structure. The final result
is required to obey the symmetry of the space group. This is also an average
structure probe--diffuse scattering intensity is ignored. Thus all Bi atomic
positions in \bto, for example, will be equivalent to each other as required
by $Fd\overline{3}m$.

Despite these constrains, the MEM affords an excellent view of average Bi
displacements. No prior description of Bi displacements, or even a knowledge
of their  existence, is used to produce them in the isosurface
neutron density unit cell displayed in Fig.\,\ref{fig:mem}(a). Arrows
point to the circular Bi density that forms a ring around the \obo\/ bond. 
Two-dimensional slices of the cell viewed along and normal to the \obo\/ 
bond in Figs.\,\ref{fig:mem}(b) and \ref{fig:mem}(c), respectively, show 
additional detail of the Bi nuclear density. The Bi ring appears hexagonal 
in Fig.\,\ref{fig:mem}(b), which is the required symmetry of the position
in $Fd\overline{3}m$. The strongest intensity of Bi points toward the in-plane
Ti in Fig.\,\ref{fig:mem}(b). 
This corresponds to a $96h$ site for Bi, not $96g$.\cite{avdeev_static_2002}
While the Bi shapes found by MEM agree with those from 
Rietveld refinements, there is no evidence of a tetrapod-shaped spread in the 
O$^\prime$ density in MEM. This shape would be allowed by symmetry.

\subsection{Least-squares PDF refinement}

\begin{figure}
\centering\includegraphics[width=8cm]{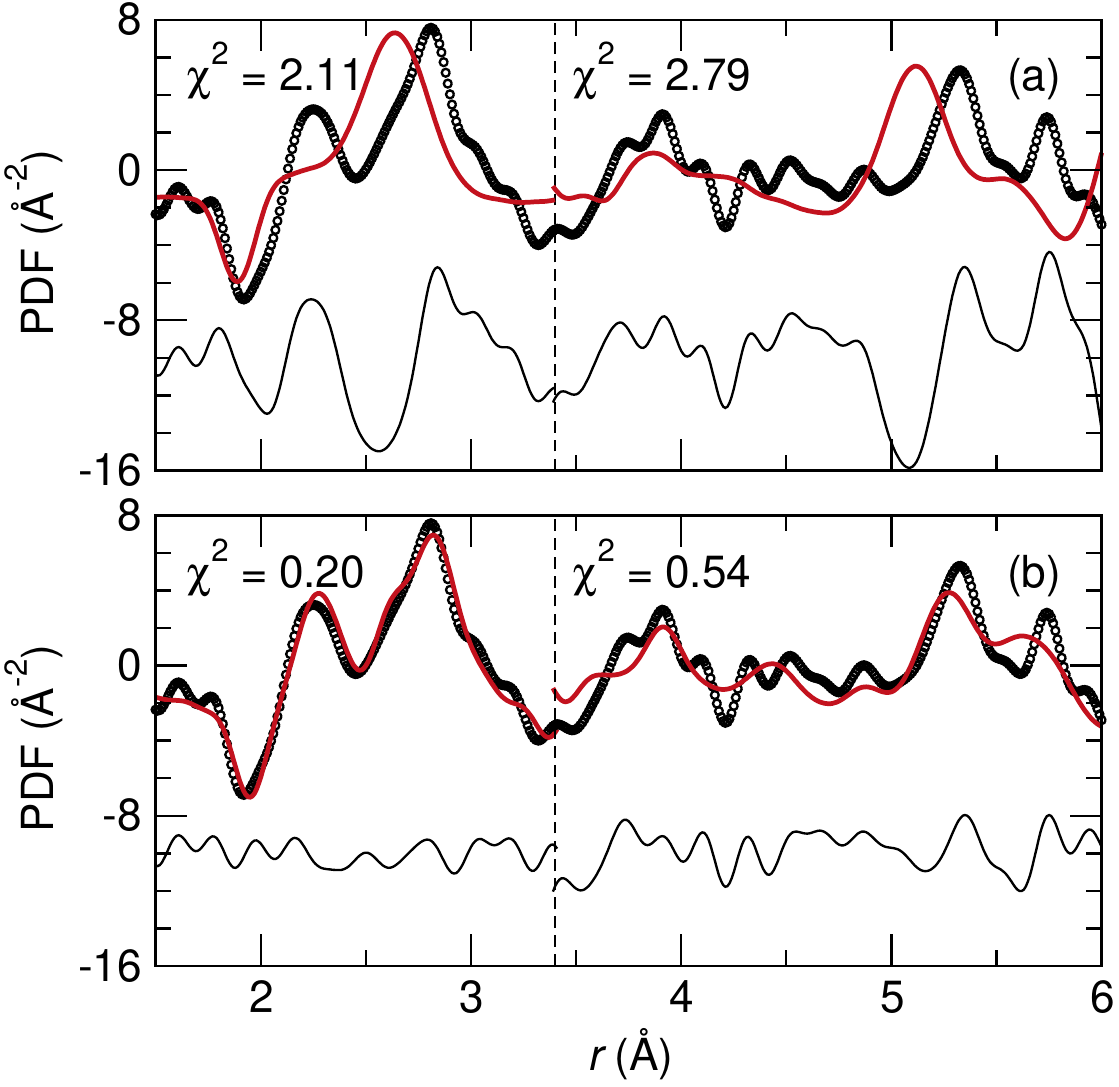}\\
\caption{(Color online) Least-squares refinements of the \bto\/ PDF at 14 K.
Panel (a) shows the fit to an ideal average structure (Bi centered on $16c$ sites with disc-shaped
thermal parameters) as shown in Fig.\,\ref{fig:unitcell}(a).
Panel (b) is the $Cm$ structure from first-principles . The $\chi^2$ is shown for two fits per
model: a fit up to 3.4 \AA\/ and a fit from 3.4 \AA\/ to 6 \AA. Difference curves are
shown below.}
\label{fig:pdfgui}
\end{figure}

\begin{figure}
\centering\includegraphics[width=8cm]{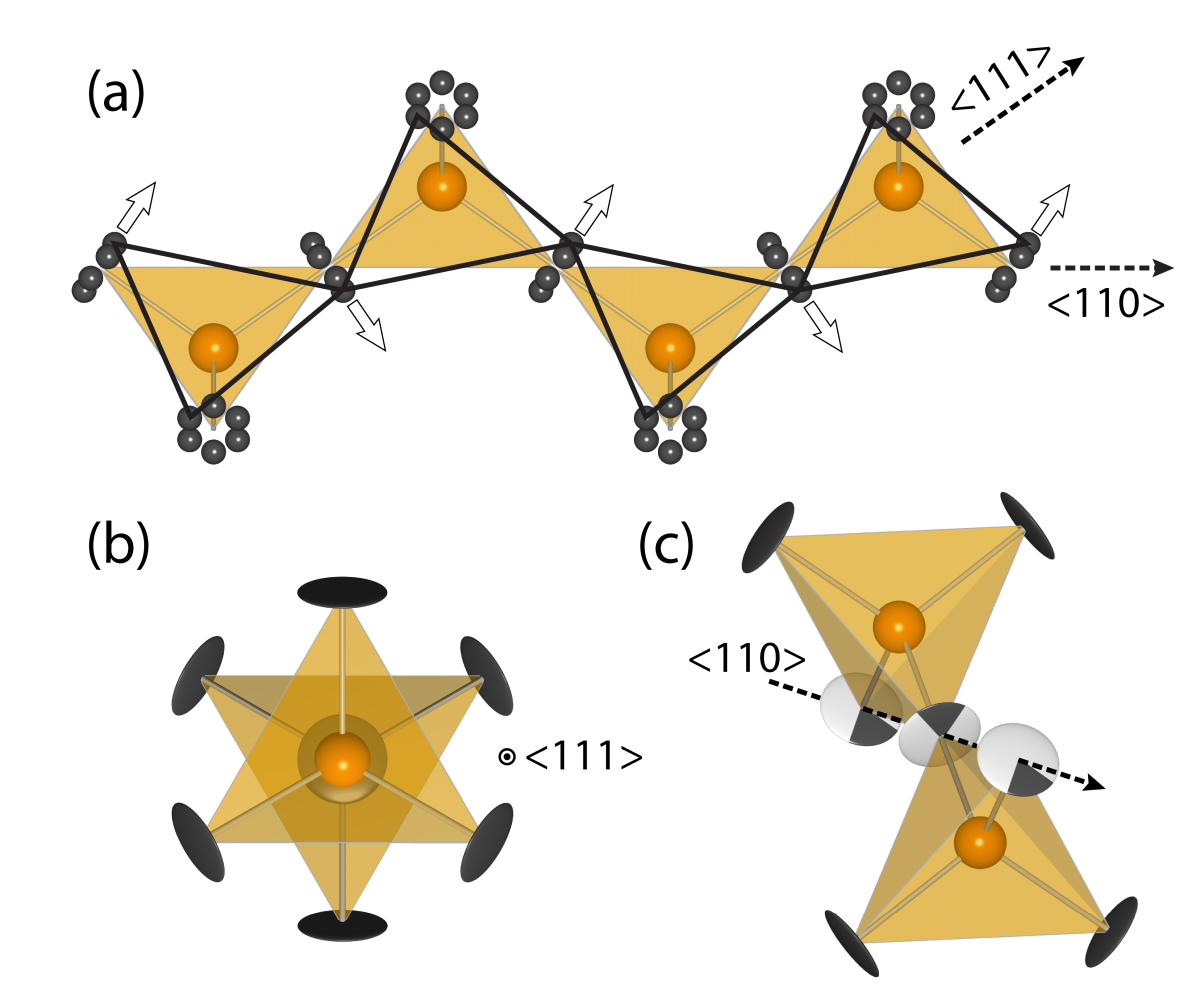} \\
\caption{(Color online) In (a), the ideal \obi\,sublattice (orange tetrahedra) 
with split-atom Bi positions (black) can cooperatively order in a zig-zag (arrowed), 
$\beta$-cristobalite-like pattern ordering seen as solid black lines. We use a 
double-tetrahedral model viewed along the O$^\prime$--O$^\prime$ direction in 
(b) to investigate local correlations. This shape is 
viewed in (c) near the Bi--Bi direction with the most-probable
wedges of the circular Bi ADPs shown schematically for
zig-zag ordering. }
\label{fig:chain}
\end{figure}

When displacements of atomic positions and their correlations have been 
proposed in the pyrochlore structure, it has been compared to the SiO$_4$ 
tetrahedral network of $\beta$-cristobalite, which is isostructural
to the \obi\/ sublattice.\cite{tabira_structured_1999,nguyen_local_2007,goodwin_real-space_2007}
This configuration would have long-range zig-zag ordering of Bi displacements shown in 
Fig.\,\ref{fig:chain}. These distortions may be ordered along any of the three 
Bi--Bi directions in the lattice, and the distortions in one direction
need not be correlated with those in another.\cite{withers_structure_1989,
tucker_dynamic_2001} We use least-squares PDF refinement (a small-box technique) to
compare the observed local structure to two models: the $Fd\bar{3}m$ Rietveld-derived
average model with
large ADPs, and the $Cm$ ordered structure from first-principles calculations.

The average model fit to the experimental PDF is shown in Fig.\,\ref{fig:pdfgui}(a).
The large, disc-shaped ADPs on the Bi positions cause the peak at 2.2 \AA\/ to flatten and
disappear,
and the fit does not significantly improve as $r$ increases. The short-range fit
up to $r = 3.4$ \AA\/ and the medium-range fit from $3.4 \leq r \leq 6$ \AA\/ 
both give high $\chi^2$ values. The polar, ordered model fit in Fig.\,\ref{fig:pdfgui}(b)
describes the low-$r$ region very well. Here, a single Bi--O$^\prime$ distance is well-defined,
so the peak at 2.2 \AA\/ appears. The medium-range fit is poorer than the short-range fit,
but still much better than the average model.
From the Rietveld, MEM, and least-squares PDF analyses we know that the structure of \bto\/
shares some attributes of the average structure (lattice parameters, averaged positions)
and the local $Cm$ model (bond distances and angles at low $r$). Analysis by
least-squares refinement of the PDF is limited because we must construct
a model that resembles both $Fd\overline{3}m$ and $Cm$. Due to the complex disorder present in
\bto, we conduct reverse Monte Carlo simulations, where large-box models are generated
by fits to the data (starting from the ideal structure), and extract statistical measures
of these models.

\subsection{Reverse Monte Carlo simulations}

RMC simulations utilize large-box modeling with periodic boundary conditions 
and are not constrained by 
symmetry.\cite{proffen_analysis_1997,tucker_application_2001,tucker_rmcprofile_2007} 
This provides two principal advantages when simulating crystalline materials: 
the ability to model nuclear positions with arbitrary shapes, and the
ability to investigate correlations between atoms on the \AA\/ length-scale. 
We profit from both when modeling \bto. First, the true shapes of Bi and 
O$^\prime$ displacements must be elucidated. Second, Bi--Bi nearest-neighbor 
correlations may lead to cooperative distortions of the \obi\/ sublattice. 
This type of short-range order would be a signature of local regions
where the lone-pair-active Bi displacements effect polar domains. 

\begin{figure}
\centering\includegraphics[width=8cm]{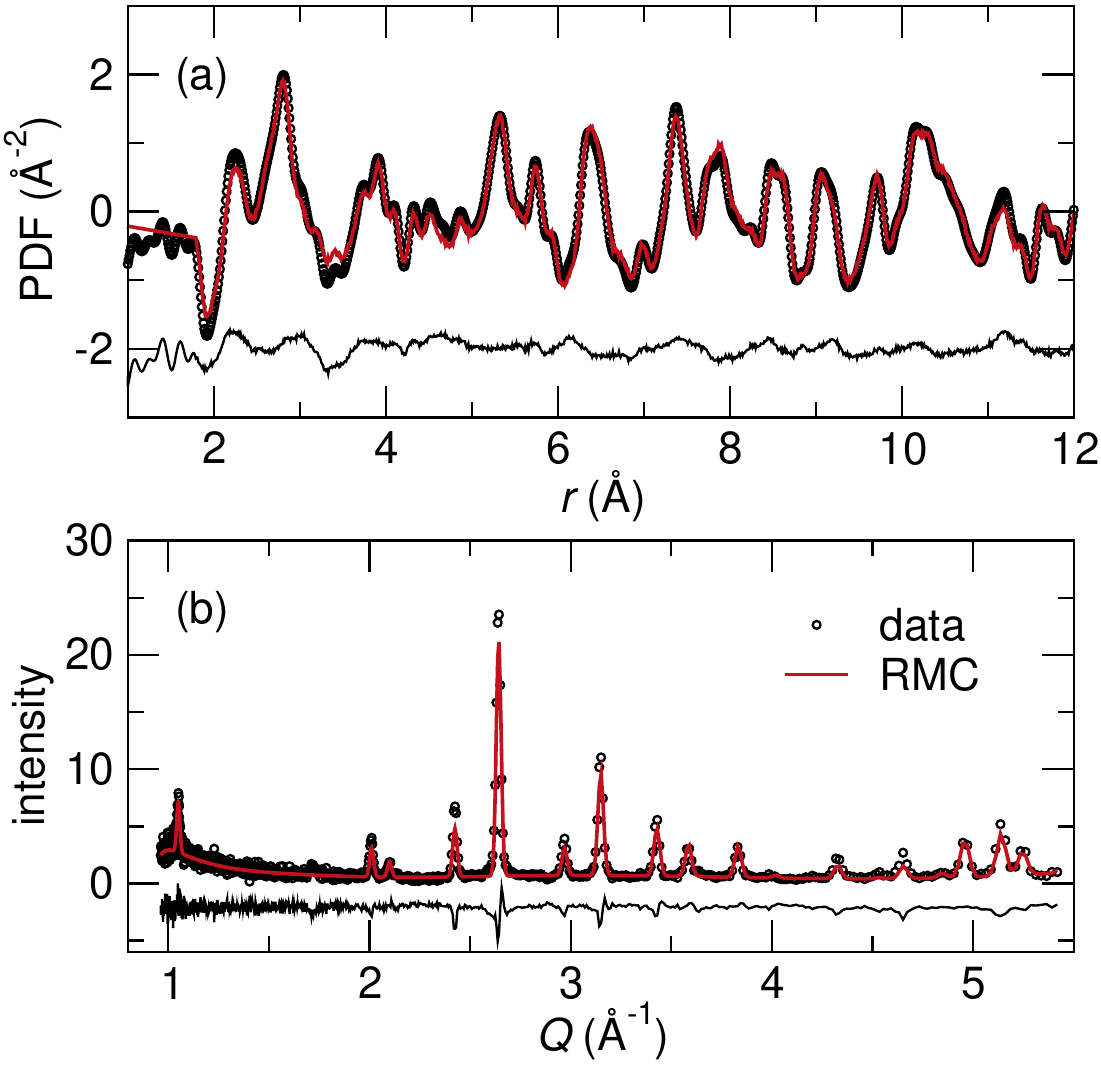}\\
\caption{(Color online) RMC fits to 14\,K \bto\/ experimental data shown for (a) the
PDF $G(r)$ and (b) the Bragg profile. The RMC simulations employ supercells 
($5 \times 5 \times 5$, $\approx 52$ \AA\/ per side) that fit both data sets 
simultaneously.}
\label{fig:rmc_fits}
\end{figure}

\subsubsection{Average atomic displacements}

The real-space local structure (coordination environments, short-range 
correlations, \textit{etc}.) of the RMC supercell are driven by the fit to the 
PDF, while the Bragg fit constrains the long-range periodicity of the 
structure (or lack thereof) and ensures reliable displacement parameters.
In the case of \bto, we monitor the agreement of Rietveld and RMC ADPs  
(calculated as the mean-square displacement from ideal positions).
RMC simulations were started with the ideal pyrochlore lattice with 
all Bi atoms on $16c$ positions. As the simulation progresses, Bi atoms are 
always observed to move off the central position to form a ring that resembles 
the Rietveld split-atom model. The fits to the 14 K PDF and Bragg profile are shown in 
Fig.\,\ref{fig:rmc_fits}(a,b).  

\begin{figure}
\centering\includegraphics[width=8cm]{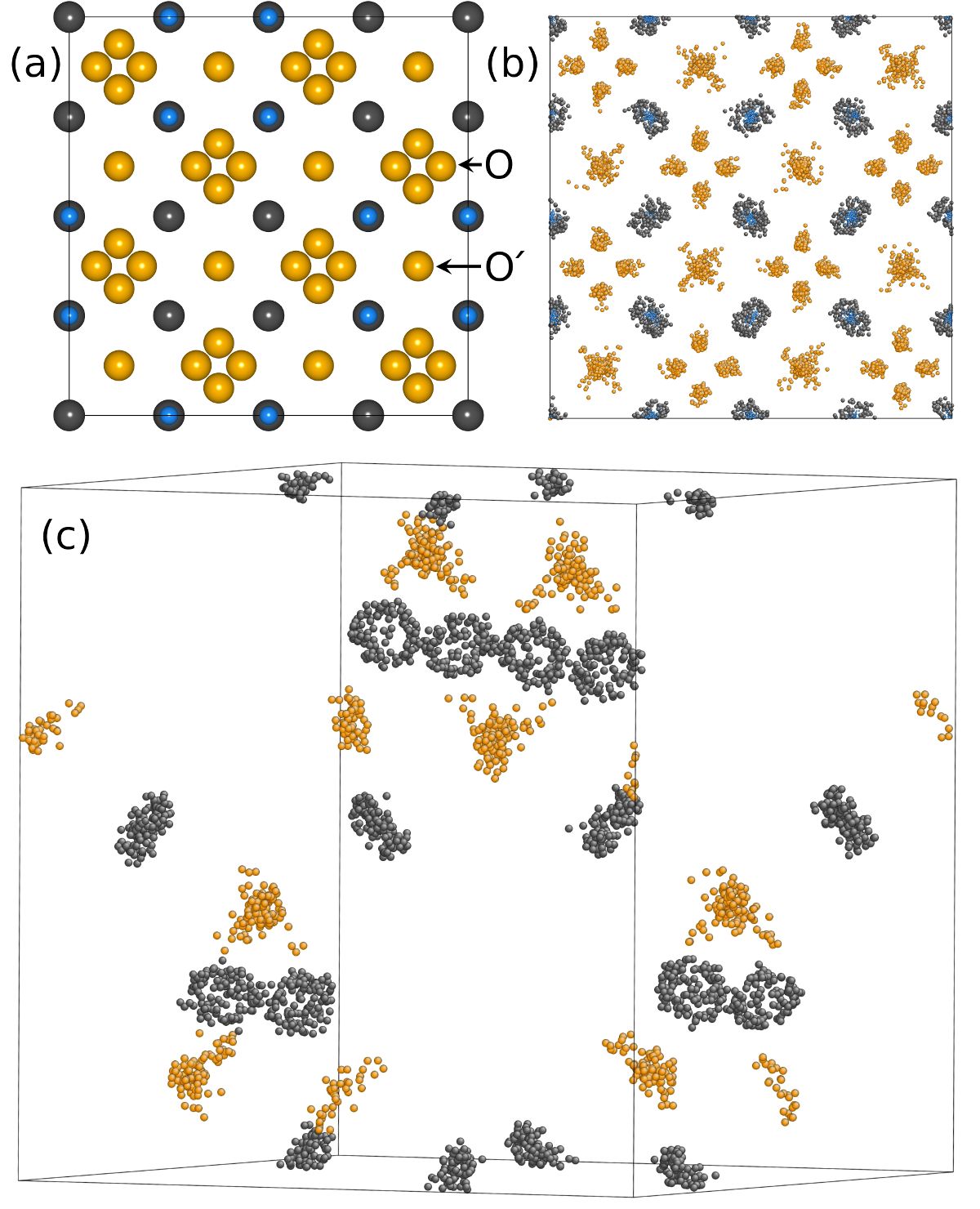}\\
\caption{(Color online) The ideal, isotropic \bto\,unit cell is plotted in (a) along the 
$a$ direction, with Bi atoms in black, Ti blue, and O orange. This unit cell 
can be visually compared with the folded 14\,K RMC supercell in (b). 
Positional disorder is much greater in Bi and O$^\prime$ (which appear 
$X$-shaped in this projection) than in Ti and O (appear as clusters of 
four O). In (c), the \obi\,sublattice is plotted near $\langle110\rangle$ to 
show the Bi rings.} \label{fig:cell_comparison}
\end{figure}

A comparison of the ideal unit cell with a folded RMC supercell is shown in
Fig.\,\ref{fig:cell_comparison}. In Fig.\,\ref{fig:cell_comparison}(b) we 
fold the 125 unit cells of a supercell into a single unit cell. The
result is a cell with ``point clouds'' on each atomic position that represent
a map of the nuclear scattering density on each site in \bto. Of particular 
note is the large spread of Bi and O$^\prime$ point clouds in 
Fig.\,\ref{fig:cell_comparison}(b) in comparison to those of Ti and O. These
distributions agree quantitatively with the ADPs from Rietveld refinement.

\begin{figure}
\centering\includegraphics[width=8cm]{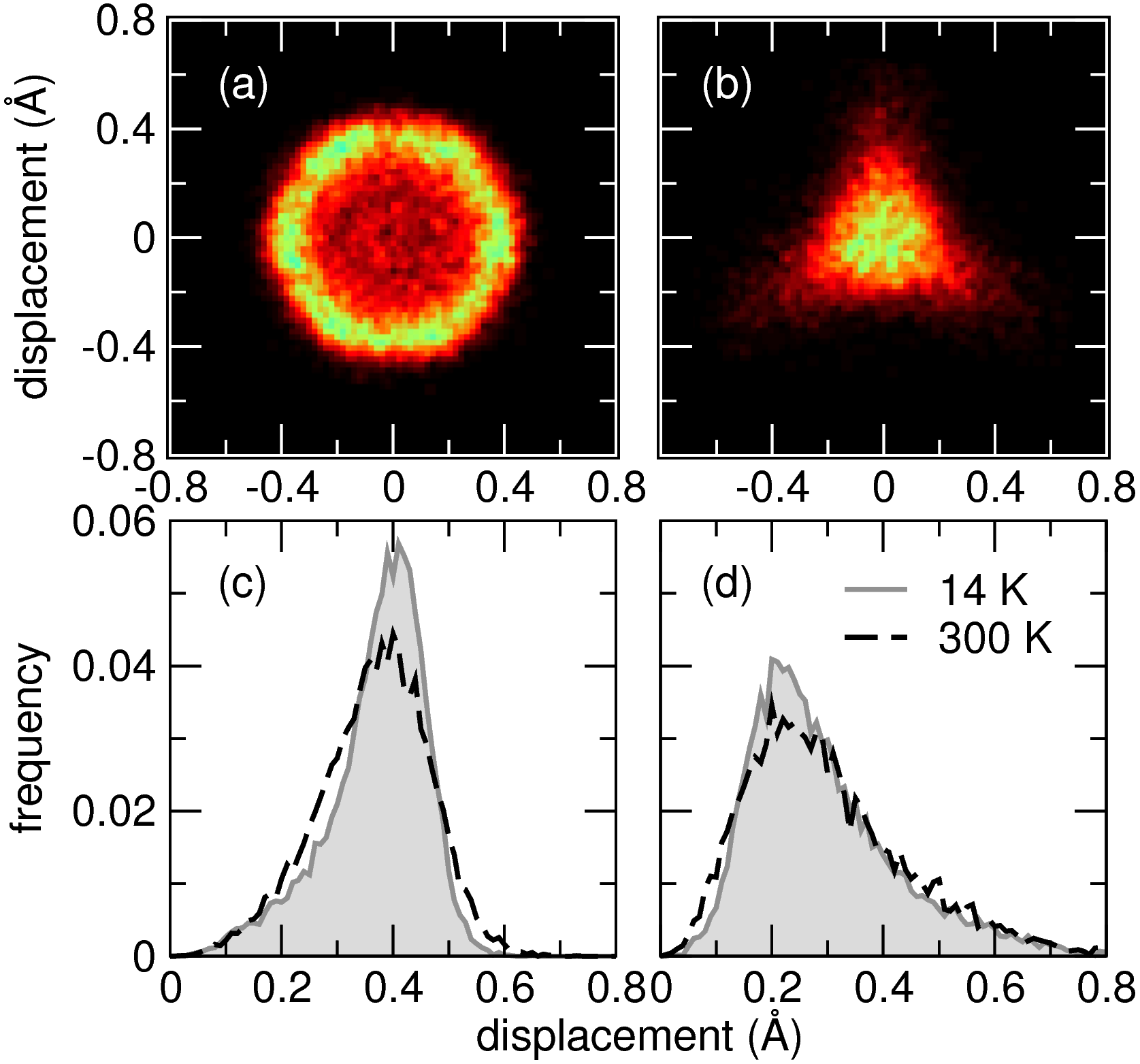}\\
\caption{(Color online) Atomic positions for (a) Bi and (b) O$^\prime$ at 14\,K from 14 
independent RMC simulations are plotted in the $\{111\}$ plane normal to 
the \obo\ bond. Histograms of displacement distances are plotted for 14 K and
300 K simulations in (c) and (d), with the 14 K histograms shaded for clarity.
The Bi median displacement of 0.4\,\AA\/ agrees with the Bi 
Rietveld ADP.}
\label{fig:rmc_shapes}
\end{figure}

\begin{figure}
\centering\includegraphics[width=8cm]{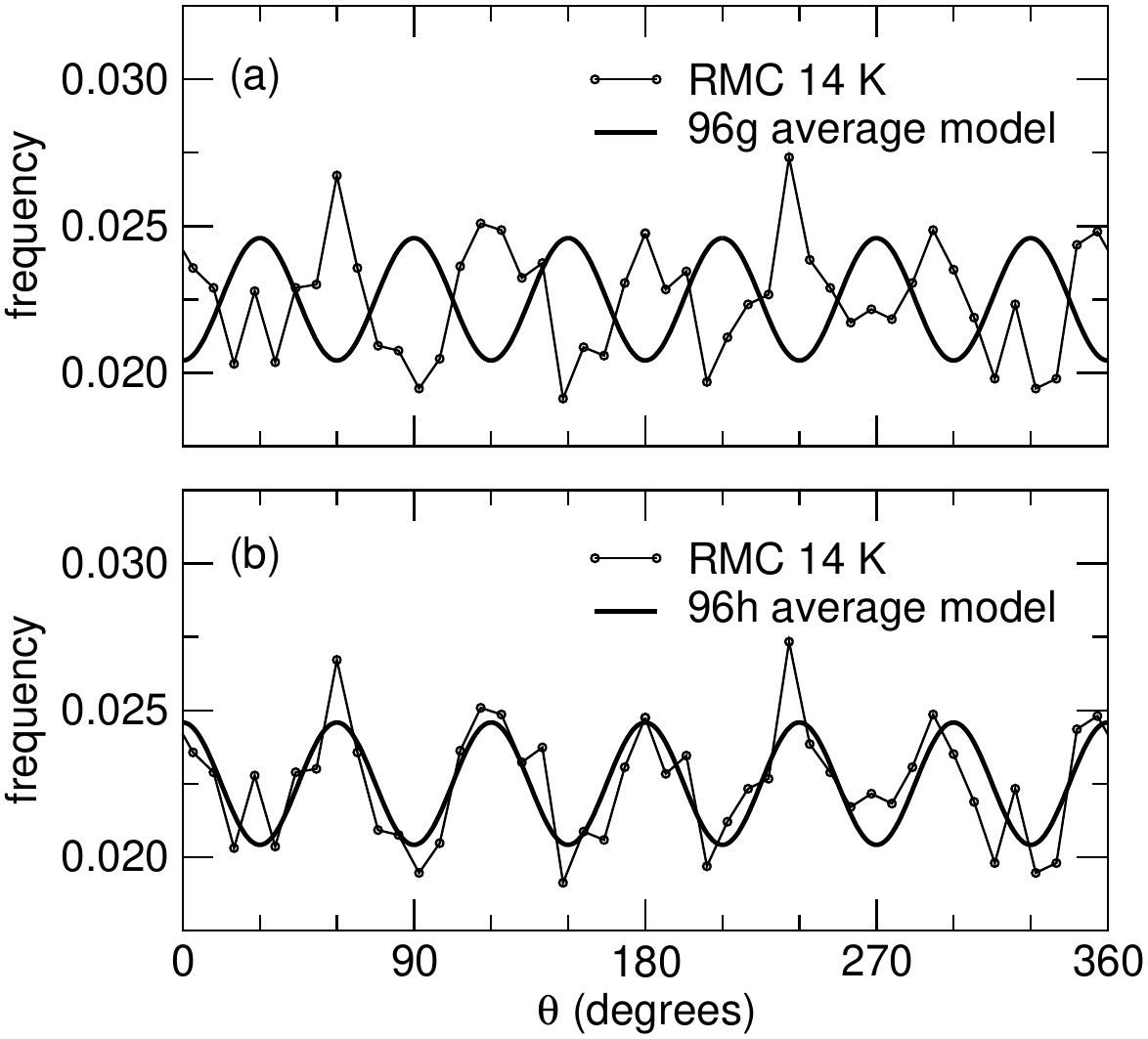}\\
\caption{Histogram of the average Bi displacement angle $\theta$ around the \obo\ bond
at 14 K. Solid lines show the expected six-fold cosine curves that indicate
preference for (a) $96g$ or (b) $96h$ hexagonal ring symmetry.}
\label{fig:bi_symmetry}
\end{figure}

The \obi\,sublattice from a 14 K RMC simulation is shown in 
Fig.\,\ref{fig:cell_comparison}(c), viewed close to the $\langle110\rangle$
direction. From this viewpoint, Bi rings are evident. These rings open normal 
to the \obo\, bonds, as expected from Rietveld ADPs and MEM. The average shape of a Bi ring is seen
in Fig.\,\ref{fig:rmc_shapes}(a). Viewed along the \obo\,direction, 
it appears circular with a radius of about 0.4\,\AA.  A histogram of Bi distances 
from the ideal $16c$ site is given in Fig.\,\ref{fig:rmc_shapes}(c), with a 
clear maximum at the ring radius. The same procedure applied to O$^\prime$ 
yields a tetrapodal object. Seen in Fig.\,\ref{fig:rmc_shapes}(b), the arms 
of this tetrapod point away from the four neighboring Bi atoms. One 
arm of the tetrapod is in the center of the plot, pointing normal to the page. 
A key distinction between O$^\prime$ and Bi is that the dense cluster of O$^\prime$
positions is still centered on the ideal $8a$ position, seen as a bright
cluster in the center of Fig.\,\ref{fig:rmc_shapes}(b). In contrast, Bi
intensity is low at the center and most intensity lies on the ring perimeter.
The displacement histograms in Figs.\,\ref{fig:rmc_shapes}(c,d) reveal similar
shapes in RMC simulations of 14 K and 300 K data, with a slight 
broadening at high temperature. This supports the idea that these displacements
are frozen at high temperatures and do not fundamentally change with cooling.

The RMC model corroborates with Rietveld and MEM while providing further evidence for a tendency toward
a sixfold Bi ring. In Fig.\,\ref{fig:bi_symmetry}(a), a histogram of the the rotation
angle $\theta$ of Bi around
the $\langle111\rangle$ axis is fit to a sixfold cosine curve. This curve, with a minimum
at $\theta = 0^\circ$, corresponds to the $96g$ sites and does not fit the data. Instead,
Fig.\,\ref{fig:bi_symmetry}(b) shows excellent agreement between the RMC result and 
the $96h$ six-fold curve, which is shifted by
30$^\circ$ from $96g$. Note however that the intensity at minima in the curves is not zero;  there
is still some tendency for Bi to lie at any $\theta$. It is not known
whether increased preference for the $96h$ site would push the minima to be nearly zero.
It is therefore possible that stronger ordering exists, but is obscured by the resolution of the data.

\subsubsection{Correlated distortions: double-tetrahedral model}

\begin{figure}
\centering\includegraphics[width=8cm]{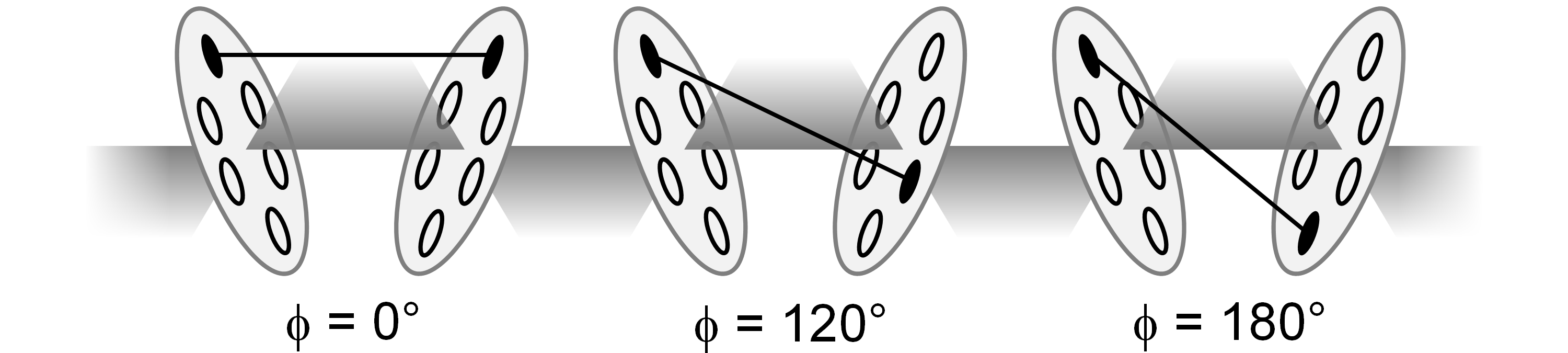}\\
\caption{Schematic of possible values of $\phi$, probing nearest-neighbor 
correlations of Bi displacements (not to scale).}
\label{fig:phi_schematic}
\end{figure}

Cooperative $\beta$-cristobalite correlations can be visualized in the 
polyhedral configuration given in Fig.\,\ref{fig:chain}.
The motif of two corner-linked \obi\,tetrahedra is the basis of
our analysis. The zig-zag pattern is comprised of Bi that are all in the same
plane (denoted with arrows). This plane is defined by the displacement
of any Bi atom in the chain, so the central Bi atom can only
participate in zig-zag ordering in one direction at a time. The two
inline nearest neighbors of the central atom should have a tendency to 
displace opposite from the central
Bi displacement vector. We can therefore quantify the correlation by examining
the angle $\phi$ between the displacement vector of the central Bi versus the displacement
vectors of the two relevant nearest neighbors. Each double-tetrahedral shape in the supercell
is examined in this manner. A schematic showing selected orientations of neighboring
Bi displacements and their corresponding values of $\phi$ is given in Fig.\,\ref{fig:phi_schematic}.

\begin{figure}
\centering\includegraphics[width=8cm]{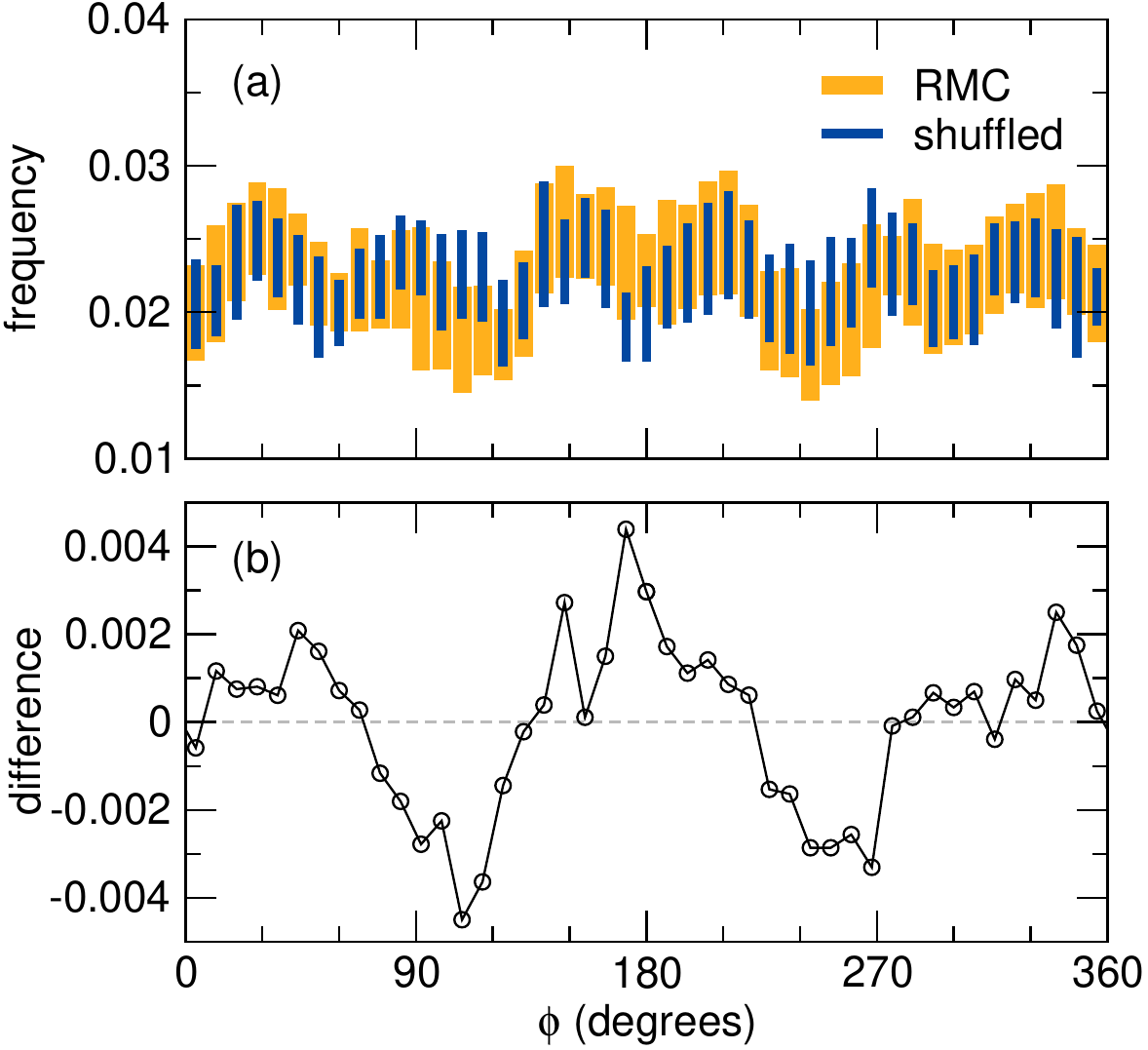}\\
\caption{(Color online) Histograms (a) of the Bi-Bi angle $\phi$ for neighboring
displacements, with error bars of $\pm \sigma$. The difference between the RMC result
and the same configurations with randomly shuffled Bi displacements is shown in (b).
The difference curve peaks at $\phi = 180^\circ$, indicating some zig-zag-like correlation.
Angles around $\phi = 120^\circ$ seem unfavorable.}
\label{fig:phi_plot}
\end{figure}

In Fig.\,\ref{fig:phi_plot}(a), the relative angle $\phi$ histogram is plotted  
for the RMC simulations. It does not have the same six-fold modulation as the averaged
Bi angles $\theta$ in Fig.\,\ref{fig:bi_symmetry}(b) because $\phi$ is defined relative
to its \emph{neighbors}, while $\theta$ is defined relative to the \emph{crystal axes}.
We investigate the tendency for ordering by shuffling Bi displacements: the
set of all individual Bi displacements  from their ideal sites is preserved, but
redistributed randomly among the Bi atoms. As a result, their orientations with respect
to each other are disrupted. Note that, as a result of the removal of this local
correlation, the shuffled $\phi$ curve is simply a six-fold cosine curve with modulation
analogous to $\theta$ in Fig.\,\ref{fig:bi_symmetry}(b). The difference between the
RMC and shuffled $\phi$, plotted
in Fig.\,\ref{fig:phi_plot}(b), shows
a peak at $\phi = 180^\circ$, indicating a preference for zig-zag ordering.
Correlations where $60^\circ < \phi < 120^\circ$ are not preferred.
While these results point to anti-alignment of Bi displacements, \bto\ should not be
considered to follow the $\beta$-cristobalite
model strictly, since Bi would need to prefer the $96g$ site rather than $96h$.

\subsubsection{Correlated distortions: continuous symmetry measures}

\begin{figure}
\centering\includegraphics[width=8cm]{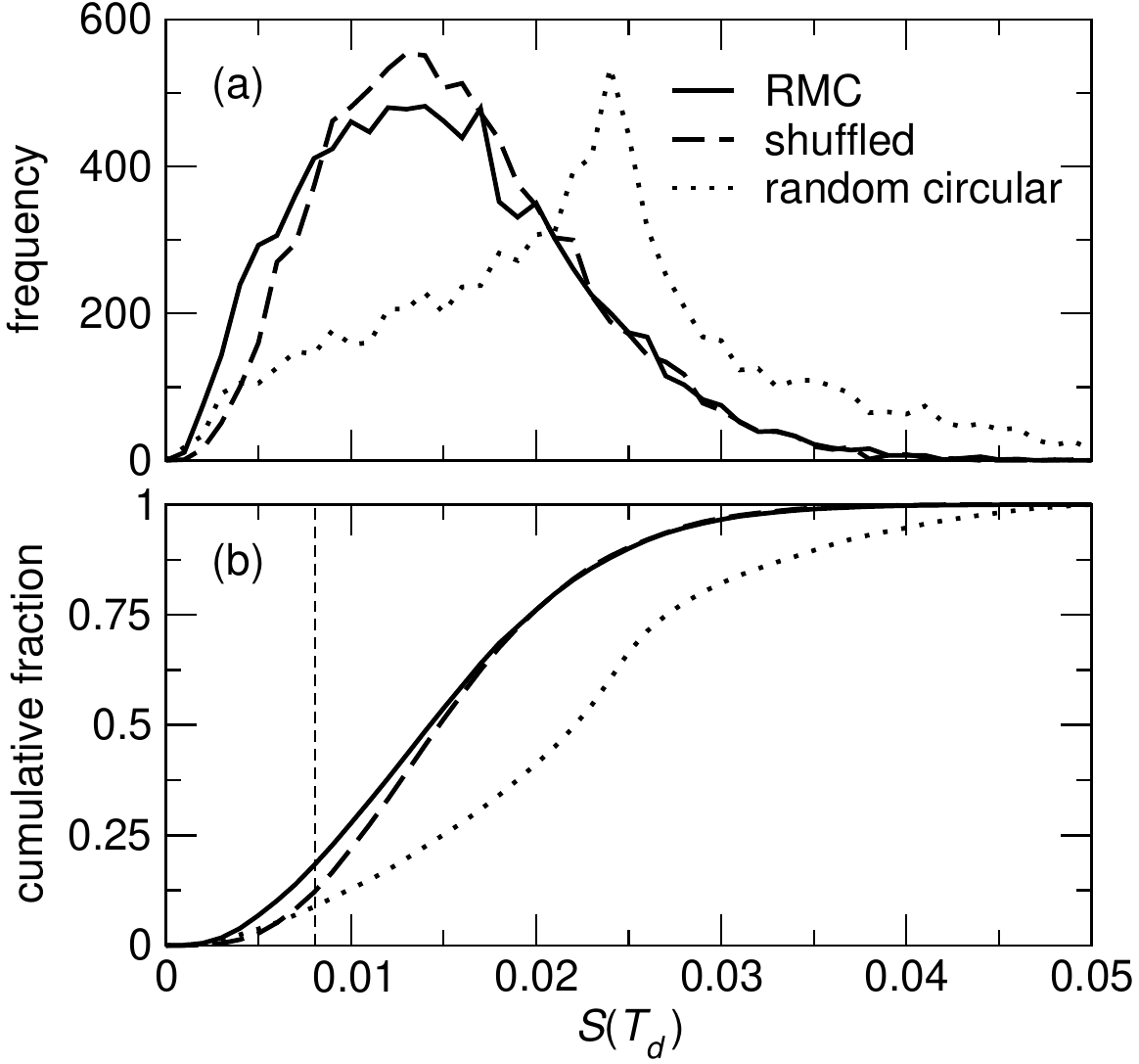}\\
\caption{Histogram (a) of the CSM tetrahedricity $S(T_d)$ for three 
supercells: the RMC supercell, the same RMC supercell with Bi displacements 
randomly swapped, and a model with random circular Bi rings of equal radius. 
The peak at higher $S(T_d)$ in the circular model indicates more
deviation from ideality than in the RMC supercells. These histograms are 
plotted as cumulative probability functions in (b), where the agreement 
between the RMC and swapped supercells suggest little evidence for cooperative 
short-range Bi ordering within tetrahedra. A fully-ordered model is denoted 
by a vertical line at $S(T_d) = 0.008$.}
\label{fig:csm}
\end{figure}

The CSM provides a quantitative measure of a polyhedron's distortion,
in the form of a ``distance'' from ideality.\cite{zabrodsky_continuous_1992,pinsky_continuous_1998,keinan_studies_2001}
In our case, a perfect 
tetrahedron would have tetrahedricity $S(T_d)$ = 0, while distortions from 
ideality increase this value toward 1. As a point of reference, the Jahn-Teller 
distorted CuO$_4$ tetrahedron in CuMn$_2$O$_4$\cite{shoemaker_unraveling_2009} 
has $S(T_d) = 0.0235$ while a square plane has $S(T_d) = \frac{1}{3}$.

Application of CSM to the \obi\/ sublattice is useful because $S(T_d)$ describes 
the correlations between four neighboring Bi atoms.  Comparison between 
models and RMC simulations are shown in Fig.\,\ref{fig:csm}. For a
model with ordered $\beta$-cristobalite-type ordering with a similar displacement 
magnitude, $S(T_d) = 0.008$. This is plotted as a vertical line 
in Fig.\,\ref{fig:csm}(b). In contrast, uncorrelated and evenly circular (with radius 0.4 \AA) Bi
displacements produce a wide distribution of $S(T_d)$, shown as a dotted line in 
Fig.\,\ref{fig:csm}(a), and again as a cumulative fraction in 
Fig.\,\ref{fig:csm}(b).

In a case where zig-zag (or otherwise correlated) distortions 
are present on most \obi\ tetrahedra in the supercell,  we expect Bi displacements
to be correlated with their nearest neighbors, and a polyhedral rigidity 
reminiscent of SiO$_4$ tetrahedra might exist. If these displacements from ideal positions were 
randomly swapped among Bi atoms, the ordering would be destroyed and $S(T_d)$ 
would increase toward the random circular model. In a case where there is no 
short-range Bi--Bi ordering, randomly swapping the 
displacements should have no effect on the $S(T_d)$ distribution.

In Fig.\,\ref{fig:csm} we compare the
$S(T_d)$ distributions for the RMC supercell and the same supercell 
with Bi displacements randomly swapped. The distributions are shown as 
histograms of $S(T_d)$ in (a) and cumulative fractions in (b). 
The random swapping slightly increases $S(T_d)$ for the RMC supercell, 
seen as a shift to the right in Fig.\,\ref{fig:csm}(b). This curve is still 
quite separated from the random circular model.

Two aspects of the Bi distribution can be gleaned from this comparison. First, 
the Bi displacements of the RMC supercell are more strongly tetrahedral 
than the random circular model. This could be due to rigidity of the \obi\
tetrahedra, or the fact that the RMC has a distribution of Bi displacements
[Fig.\,\ref{fig:rmc_shapes}(c)] whereas the circular model has strictly $r = 0.4$ \AA.
The separation between RMC and shuffled CSM points to rigidity
that may be explained by correlated Bi displacements. Just as shuffling Bi
displacements disrupts the ordering between
\emph{pairs} of Bi as viewed by $\phi$ in Fig.\,\ref{fig:phi_plot}(b), it also disrupts
ordering between sets of \emph{four} Bi displacements as viewed by CSM in Fig.\,\ref{fig:csm}(b).

\section{Conclusions} 

We have shown that a structural investigation of the strongly disordered pyrochlore \bto\/
benefits from modeling real- and reciprocal space structure simultaneously.
The combination of Rietveld refinement, MEM, least-squares PDF refinement, and RMC
simulations provide a description of the structure over many length scales. Of particular
interest is the behavior of the displaced lone-pair active Bi$^{3+}$ cation. 
Average 
structure models indicate a preference for static displacements. MEM and RMC show that
Bi prefers the $96h$ (pointing between nearby O from the TiO$_6$ sublattice)
rather than $96g$ sites (pointing toward neighboring O). Statistical measures of these
displacements are extracted from RMC supercells.

RMC simulations show that Bi displacements lie at all angles $\theta$ in a ring
normal to the \obo\ bond with radius $r \sim 0.4$\,\AA, but there is a preference
for $\theta$ corresponding to $96h$ positions. O$^\prime$ atoms are centered on the
ideal site, but in a tetrapodal shape with arms pointing away from neighboring Bi.
Comparing the RMC result with an identical supercell with shuffled Bi displacements
reveals that neighboring Bi displacements show some correlation with each other,
with a preference for $phi = 180^\circ$ alignment. This corresponds to a zig-zag type
ordering that merits further investigation.
Measures of the tetrahedricity $S(T_d)$ of the \obi\/ sublattice indicate that neighboring
displacements are weakly correlated, thus the sublattice may exhibit some rigidity.

\section{Acknowledgments}
We thank M. Pinsky and D. Avnir for providing the CSM script, and we thank K. L. Page,
R. L. Withers, M. G. Tucker, and A. L. Goodwin for helpful discussions.
This work was supported by the 
Institute for Multiscale Materials Studies and the National Science Foundation through
a Career Award (DMR 0449354) to RS and MRSEC facilities 
(DMR 0520415). Neutron scattering was performed on NPDF at the Lujan Center at 
the Los Alamos Neutron Science Center, funded by the DOE Office of Basic 
Energy Sciences. Los Alamos National Laboratory is operated by Los Alamos 
National Security, LLC under DOE Contract DE-AC52-06NA25396.

\bibliography{bto7}

\begin{thebibliography}{58}
\expandafter\ifx\csname natexlab\endcsname\relax\def\natexlab#1{#1}\fi
\expandafter\ifx\csname bibnamefont\endcsname\relax
  \def\bibnamefont#1{#1}\fi
\expandafter\ifx\csname bibfnamefont\endcsname\relax
  \def\bibfnamefont#1{#1}\fi
\expandafter\ifx\csname citenamefont\endcsname\relax
  \def\citenamefont#1{#1}\fi
\expandafter\ifx\csname url\endcsname\relax
  \def\url#1{\texttt{#1}}\fi
\expandafter\ifx\csname urlprefix\endcsname\relax\def\urlprefix{URL }\fi
\providecommand{\bibinfo}[2]{#2}
\providecommand{\eprint}[2][]{\url{#2}}

\bibitem[{\citenamefont{Ramirez}(1994)}]{ramirez_strongly_1994}
\bibinfo{author}{\bibfnamefont{A.~P.} \bibnamefont{Ramirez}},
  \bibinfo{journal}{Ann. Rev. Mater. Sci.} \textbf{\bibinfo{volume}{24}},
  \bibinfo{pages}{453} (\bibinfo{year}{1994}).

\bibitem[{\citenamefont{Ramirez et~al.}(1999)\citenamefont{Ramirez, Hayashi,
  Cava, Siddharthan, and Shastry}}]{ramirez_zero-point_1999}
\bibinfo{author}{\bibfnamefont{A.~P.} \bibnamefont{Ramirez}},
  \bibinfo{author}{\bibfnamefont{A.}~\bibnamefont{Hayashi}},
  \bibinfo{author}{\bibfnamefont{R.~J.} \bibnamefont{Cava}},
  \bibinfo{author}{\bibfnamefont{R.}~\bibnamefont{Siddharthan}},
  \bibnamefont{and} \bibinfo{author}{\bibfnamefont{B.~S.}
  \bibnamefont{Shastry}}, \bibinfo{journal}{Nature}
  \textbf{\bibinfo{volume}{399}}, \bibinfo{pages}{333} (\bibinfo{year}{1999}).

\bibitem[{\citenamefont{Morris et~al.}(2009)\citenamefont{Morris, Tennant,
  Grigera, Klemke, Castelnovo, Moessner, Czternasty, Meissner, Rule, Hoffmann
  et~al.}}]{morris_dirac_2009}
\bibinfo{author}{\bibfnamefont{D.~J.~P.} \bibnamefont{Morris}},
  \bibinfo{author}{\bibfnamefont{D.~A.} \bibnamefont{Tennant}},
  \bibinfo{author}{\bibfnamefont{S.~A.} \bibnamefont{Grigera}},
  \bibinfo{author}{\bibfnamefont{B.}~\bibnamefont{Klemke}},
  \bibinfo{author}{\bibfnamefont{C.}~\bibnamefont{Castelnovo}},
  \bibinfo{author}{\bibfnamefont{R.}~\bibnamefont{Moessner}},
  \bibinfo{author}{\bibfnamefont{C.}~\bibnamefont{Czternasty}},
  \bibinfo{author}{\bibfnamefont{M.}~\bibnamefont{Meissner}},
  \bibinfo{author}{\bibfnamefont{K.~C.} \bibnamefont{Rule}},
  \bibinfo{author}{\bibfnamefont{J.}~\bibnamefont{Hoffmann}},
  \bibnamefont{et~al.}, \bibinfo{journal}{Science}
  \textbf{\bibinfo{volume}{326}}, \bibinfo{pages}{411} (\bibinfo{year}{2009}).

\bibitem[{\citenamefont{Kadowaki et~al.}(2009)\citenamefont{Kadowaki, Doi,
  Aoki, Tabata, Sato, Lynn, Matsuhira, and Hiroi}}]{kadowaki_observation_2009}
\bibinfo{author}{\bibfnamefont{H.}~\bibnamefont{Kadowaki}},
  \bibinfo{author}{\bibfnamefont{N.}~\bibnamefont{Doi}},
  \bibinfo{author}{\bibfnamefont{Y.}~\bibnamefont{Aoki}},
  \bibinfo{author}{\bibfnamefont{Y.}~\bibnamefont{Tabata}},
  \bibinfo{author}{\bibfnamefont{T.~J.} \bibnamefont{Sato}},
  \bibinfo{author}{\bibfnamefont{J.~W.} \bibnamefont{Lynn}},
  \bibinfo{author}{\bibfnamefont{K.}~\bibnamefont{Matsuhira}},
  \bibnamefont{and} \bibinfo{author}{\bibfnamefont{Z.}~\bibnamefont{Hiroi}},
  \bibinfo{journal}{J. Phys. Soc. Japan} \textbf{\bibinfo{volume}{78}},
  \bibinfo{pages}{103706} (\bibinfo{year}{2009}).

\bibitem[{\citenamefont{Yonezawa et~al.}(2004)\citenamefont{Yonezawa, Muraoka,
  and Hiroi}}]{yonezawa_new_2004}
\bibinfo{author}{\bibfnamefont{S.}~\bibnamefont{Yonezawa}},
  \bibinfo{author}{\bibfnamefont{Y.}~\bibnamefont{Muraoka}}, \bibnamefont{and}
  \bibinfo{author}{\bibfnamefont{Z.}~\bibnamefont{Hiroi}}, \bibinfo{journal}{J.
  Phys. Soc. Japan} \textbf{\bibinfo{volume}{73}}, \bibinfo{pages}{1655}
  (\bibinfo{year}{2004}).

\bibitem[{\citenamefont{Kendziora et~al.}(2005)\citenamefont{Kendziora,
  Sergienko, Jin, He, Keppens, Sales, and Mandrus}}]{kendziora_goldstone_2005}
\bibinfo{author}{\bibfnamefont{C.~A.} \bibnamefont{Kendziora}},
  \bibinfo{author}{\bibfnamefont{I.~A.} \bibnamefont{Sergienko}},
  \bibinfo{author}{\bibfnamefont{R.}~\bibnamefont{Jin}},
  \bibinfo{author}{\bibfnamefont{J.}~\bibnamefont{He}},
  \bibinfo{author}{\bibfnamefont{V.}~\bibnamefont{Keppens}},
  \bibinfo{author}{\bibfnamefont{B.~C.} \bibnamefont{Sales}}, \bibnamefont{and}
  \bibinfo{author}{\bibfnamefont{D.}~\bibnamefont{Mandrus}},
  \bibinfo{journal}{Phys. Rev. Lett.} \textbf{\bibinfo{volume}{95}},
  \bibinfo{pages}{125503} (\bibinfo{year}{2005}).

\bibitem[{\citenamefont{Onoda and Nagaosa}(2003)}]{onoda_spin_2003}
\bibinfo{author}{\bibfnamefont{S.}~\bibnamefont{Onoda}} \bibnamefont{and}
  \bibinfo{author}{\bibfnamefont{N.}~\bibnamefont{Nagaosa}},
  \bibinfo{journal}{Phys. Rev. Lett.} \textbf{\bibinfo{volume}{90}},
  \bibinfo{pages}{196602} (\bibinfo{year}{2003}).

\bibitem[{\citenamefont{Seshadri}(2006)}]{seshadri_lone_2006}
\bibinfo{author}{\bibfnamefont{R.}~\bibnamefont{Seshadri}},
  \bibinfo{journal}{Solid State Sci.} \textbf{\bibinfo{volume}{8}},
  \bibinfo{pages}{259} (\bibinfo{year}{2006}).

\bibitem[{\citenamefont{Hector and Wiggin}(2004)}]{hector_synthesis_2004}
\bibinfo{author}{\bibfnamefont{A.~L.} \bibnamefont{Hector}} \bibnamefont{and}
  \bibinfo{author}{\bibfnamefont{S.~B.} \bibnamefont{Wiggin}},
  \bibinfo{journal}{J. Solid State Chem.} \textbf{\bibinfo{volume}{177}},
  \bibinfo{pages}{139} (\bibinfo{year}{2004}), ISSN \bibinfo{issn}{0022-4596}.

\bibitem[{\citenamefont{Hill}(2002)}]{hill_density_2002}
\bibinfo{author}{\bibfnamefont{N.~A.} \bibnamefont{Hill}},
  \bibinfo{journal}{Ann. Rev. Mater. Res.} \textbf{\bibinfo{volume}{32}},
  \bibinfo{pages}{1} (\bibinfo{year}{2002}).

\bibitem[{\citenamefont{Melot et~al.}(2009)\citenamefont{Melot, Tackett,
  {O'Brien}, Hector, Lawes, Seshadri, and Ramirez}}]{melot_large_2009}
\bibinfo{author}{\bibfnamefont{B.~C.} \bibnamefont{Melot}},
  \bibinfo{author}{\bibfnamefont{R.}~\bibnamefont{Tackett}},
  \bibinfo{author}{\bibfnamefont{J.}~\bibnamefont{{O'Brien}}},
  \bibinfo{author}{\bibfnamefont{A.~L.} \bibnamefont{Hector}},
  \bibinfo{author}{\bibfnamefont{G.}~\bibnamefont{Lawes}},
  \bibinfo{author}{\bibfnamefont{R.}~\bibnamefont{Seshadri}}, \bibnamefont{and}
  \bibinfo{author}{\bibfnamefont{A.~P.} \bibnamefont{Ramirez}},
  \bibinfo{journal}{Phys. Rev. B} \textbf{\bibinfo{volume}{79}},
  \bibinfo{pages}{224111} (\bibinfo{year}{2009}).

\bibitem[{\citenamefont{Fennie et~al.}(2007)\citenamefont{Fennie, Seshadri, and
  Rabe}}]{fennie_lattice_2007}
\bibinfo{author}{\bibfnamefont{C.~J.} \bibnamefont{Fennie}},
  \bibinfo{author}{\bibfnamefont{R.}~\bibnamefont{Seshadri}}, \bibnamefont{and}
  \bibinfo{author}{\bibfnamefont{K.~M.} \bibnamefont{Rabe}},
  \bibinfo{journal}{0712.1846}  (\bibinfo{year}{2007}),
  \urlprefix\url{http://arxiv.org/abs/0712.1846}.

\bibitem[{\citenamefont{Withers et~al.}(2004)\citenamefont{Withers, Welberry,
  Larsson, Liu, Norén, Rundlöf, and Brink}}]{withers_local_2004}
\bibinfo{author}{\bibfnamefont{R.~L.} \bibnamefont{Withers}},
  \bibinfo{author}{\bibfnamefont{T.~R.} \bibnamefont{Welberry}},
  \bibinfo{author}{\bibfnamefont{A.~K.} \bibnamefont{Larsson}},
  \bibinfo{author}{\bibfnamefont{Y.}~\bibnamefont{Liu}},
  \bibinfo{author}{\bibfnamefont{L.}~\bibnamefont{Norén}},
  \bibinfo{author}{\bibfnamefont{H.}~\bibnamefont{Rundlöf}}, \bibnamefont{and}
  \bibinfo{author}{\bibfnamefont{F.~J.} \bibnamefont{Brink}},
  \bibinfo{journal}{J. Solid State Chem.} \textbf{\bibinfo{volume}{177}},
  \bibinfo{pages}{231} (\bibinfo{year}{2004}), ISSN \bibinfo{issn}{0022-4596}.

\bibitem[{\citenamefont{Goodwin et~al.}(2007)\citenamefont{Goodwin, Withers,
  and Nguyen}}]{goodwin_real-space_2007}
\bibinfo{author}{\bibfnamefont{A.~L.} \bibnamefont{Goodwin}},
  \bibinfo{author}{\bibfnamefont{R.~L.} \bibnamefont{Withers}},
  \bibnamefont{and} \bibinfo{author}{\bibfnamefont{H.}~\bibnamefont{Nguyen}},
  \bibinfo{journal}{J. Phys. Cond. Mat.} \textbf{\bibinfo{volume}{19}},
  \bibinfo{pages}{335216} (\bibinfo{year}{2007}).

\bibitem[{\citenamefont{Liu et~al.}(2009)\citenamefont{Liu, Withers, Nguyen,
  Elliott, Ren, and Chen}}]{liu_displacive_2009}
\bibinfo{author}{\bibfnamefont{Y.}~\bibnamefont{Liu}},
  \bibinfo{author}{\bibfnamefont{R.~L.} \bibnamefont{Withers}},
  \bibinfo{author}{\bibfnamefont{H.~B.} \bibnamefont{Nguyen}},
  \bibinfo{author}{\bibfnamefont{K.}~\bibnamefont{Elliott}},
  \bibinfo{author}{\bibfnamefont{Q.}~\bibnamefont{Ren}}, \bibnamefont{and}
  \bibinfo{author}{\bibfnamefont{Z.}~\bibnamefont{Chen}}, \bibinfo{journal}{J.
  Solid State Chem.} \textbf{\bibinfo{volume}{182}}, \bibinfo{pages}{2748}
  (\bibinfo{year}{2009}).

\bibitem[{\citenamefont{Tabira et~al.}(2001)\citenamefont{Tabira, Withers,
  Yamada, and Ishizawa}}]{tabira_annular_2001}
\bibinfo{author}{\bibfnamefont{Y.}~\bibnamefont{Tabira}},
  \bibinfo{author}{\bibfnamefont{R.~L.} \bibnamefont{Withers}},
  \bibinfo{author}{\bibfnamefont{T.}~\bibnamefont{Yamada}}, \bibnamefont{and}
  \bibinfo{author}{\bibfnamefont{N.}~\bibnamefont{Ishizawa}},
  \bibinfo{journal}{Z. Kristallogr.} \textbf{\bibinfo{volume}{216}},
  \bibinfo{pages}{92} (\bibinfo{year}{2001}).

\bibitem[{\citenamefont{Egami and Billinge}(2003)}]{egami_underneath_2003}
\bibinfo{author}{\bibfnamefont{T.}~\bibnamefont{Egami}} \bibnamefont{and}
  \bibinfo{author}{\bibfnamefont{S.~J.~L.} \bibnamefont{Billinge}},
  \emph{\bibinfo{title}{Underneath the Bragg Peaks, Volume 7: Structural
  Analysis of Complex Materials}} (\bibinfo{publisher}{Pergamon},
  \bibinfo{year}{2003}), ISBN \bibinfo{isbn}{0080426980}.

\bibitem[{\citenamefont{Proffen}(2000)}]{proffen_analysis_2000}
\bibinfo{author}{\bibfnamefont{T.}~\bibnamefont{Proffen}},
  \bibinfo{journal}{cond-mat/0002388}  (\bibinfo{year}{2000}),
  \urlprefix\url{http://arxiv.org/abs/cond-mat/0002388}.

\bibitem[{\citenamefont{{McGreevy} and Pusztai}(1988)}]{mcgreevy_reverse_1988}
\bibinfo{author}{\bibfnamefont{R.~L.} \bibnamefont{{McGreevy}}}
  \bibnamefont{and} \bibinfo{author}{\bibfnamefont{L.}~\bibnamefont{Pusztai}},
  \bibinfo{journal}{Mol. Simulat.} \textbf{\bibinfo{volume}{1}},
  \bibinfo{pages}{359} (\bibinfo{year}{1988}).

\bibitem[{\citenamefont{{McGreevy}}(2001)}]{mcgreevy_reverse_2001}
\bibinfo{author}{\bibfnamefont{R.~L.} \bibnamefont{{McGreevy}}},
  \bibinfo{journal}{J. Phys. Cond. Matt.} \textbf{\bibinfo{volume}{13}},
  \bibinfo{pages}{R877} (\bibinfo{year}{2001}).

\bibitem[{\citenamefont{Tucker et~al.}(2001{\natexlab{a}})\citenamefont{Tucker,
  Dove, and Keen}}]{tucker_application_2001}
\bibinfo{author}{\bibfnamefont{M.~G.} \bibnamefont{Tucker}},
  \bibinfo{author}{\bibfnamefont{M.~T.} \bibnamefont{Dove}}, \bibnamefont{and}
  \bibinfo{author}{\bibfnamefont{D.~A.} \bibnamefont{Keen}},
  \bibinfo{journal}{J. Appl. Cryst.} \textbf{\bibinfo{volume}{34}},
  \bibinfo{pages}{630} (\bibinfo{year}{2001}{\natexlab{a}}).

\bibitem[{\citenamefont{Sartbaeva et~al.}(2007)\citenamefont{Sartbaeva, Wells,
  Thorpe, Bozin, and Billinge}}]{sartbaeva_quadrupolar_2007}
\bibinfo{author}{\bibfnamefont{A.}~\bibnamefont{Sartbaeva}},
  \bibinfo{author}{\bibfnamefont{S.~A.} \bibnamefont{Wells}},
  \bibinfo{author}{\bibfnamefont{M.~F.} \bibnamefont{Thorpe}},
  \bibinfo{author}{\bibfnamefont{E.~S.} \bibnamefont{Bozin}}, \bibnamefont{and}
  \bibinfo{author}{\bibfnamefont{S.~J.~L.} \bibnamefont{Billinge}},
  \bibinfo{journal}{Phys. Rev. Lett.} \textbf{\bibinfo{volume}{99}},
  \bibinfo{pages}{155503} (\bibinfo{year}{2007}).

\bibitem[{\citenamefont{Tucker et~al.}(2001{\natexlab{b}})\citenamefont{Tucker,
  Squires, Dove, and Keen}}]{tucker_dynamic_2001}
\bibinfo{author}{\bibfnamefont{M.~G.} \bibnamefont{Tucker}},
  \bibinfo{author}{\bibfnamefont{M.~P.} \bibnamefont{Squires}},
  \bibinfo{author}{\bibfnamefont{M.~T.} \bibnamefont{Dove}}, \bibnamefont{and}
  \bibinfo{author}{\bibfnamefont{D.~A.} \bibnamefont{Keen}},
  \bibinfo{journal}{J. Phys. Cond. Mat.} \textbf{\bibinfo{volume}{13}},
  \bibinfo{pages}{403} (\bibinfo{year}{2001}{\natexlab{b}}).

\bibitem[{\citenamefont{Adams and Swenson}(2005)}]{adams_bond_2005}
\bibinfo{author}{\bibfnamefont{S.}~\bibnamefont{Adams}} \bibnamefont{and}
  \bibinfo{author}{\bibfnamefont{J.}~\bibnamefont{Swenson}},
  \bibinfo{journal}{J. Phys. Cond. Matt.} \textbf{\bibinfo{volume}{17}},
  \bibinfo{pages}{S87} (\bibinfo{year}{2005}).

\bibitem[{\citenamefont{Shoemaker et~al.}(2009)\citenamefont{Shoemaker, Li, and
  Seshadri}}]{shoemaker_unraveling_2009}
\bibinfo{author}{\bibfnamefont{D.~P.} \bibnamefont{Shoemaker}},
  \bibinfo{author}{\bibfnamefont{J.}~\bibnamefont{Li}}, \bibnamefont{and}
  \bibinfo{author}{\bibfnamefont{R.}~\bibnamefont{Seshadri}},
  \bibinfo{journal}{J. Am. Chem. Soc.} \textbf{\bibinfo{volume}{131}},
  \bibinfo{pages}{11450} (\bibinfo{year}{2009}).

\bibitem[{\citenamefont{Zabrodsky et~al.}(1992)\citenamefont{Zabrodsky, Peleg,
  and Avnir}}]{zabrodsky_continuous_1992}
\bibinfo{author}{\bibfnamefont{H.}~\bibnamefont{Zabrodsky}},
  \bibinfo{author}{\bibfnamefont{S.}~\bibnamefont{Peleg}}, \bibnamefont{and}
  \bibinfo{author}{\bibfnamefont{D.}~\bibnamefont{Avnir}}, \bibinfo{journal}{J.
  Am. Chem. Soc.} \textbf{\bibinfo{volume}{114}}, \bibinfo{pages}{7843}
  (\bibinfo{year}{1992}).

\bibitem[{\citenamefont{Pinsky and Avnir}(1998)}]{pinsky_continuous_1998}
\bibinfo{author}{\bibfnamefont{M.}~\bibnamefont{Pinsky}} \bibnamefont{and}
  \bibinfo{author}{\bibfnamefont{D.}~\bibnamefont{Avnir}},
  \bibinfo{journal}{Inorg. Chem.} \textbf{\bibinfo{volume}{37}},
  \bibinfo{pages}{5575} (\bibinfo{year}{1998}).

\bibitem[{\citenamefont{Keinan and Avnir}(2001)}]{keinan_studies_2001}
\bibinfo{author}{\bibfnamefont{S.}~\bibnamefont{Keinan}} \bibnamefont{and}
  \bibinfo{author}{\bibfnamefont{D.}~\bibnamefont{Avnir}}, \bibinfo{journal}{J.
  Che. Soc., Dalton Trans.} pp. \bibinfo{pages}{941--947}
  (\bibinfo{year}{2001}).

\bibitem[{\citenamefont{{Yogev-Einot} and
  Avnir}(2004)}]{yogev-einot_pressure_2004}
\bibinfo{author}{\bibfnamefont{D.}~\bibnamefont{{Yogev-Einot}}}
  \bibnamefont{and} \bibinfo{author}{\bibfnamefont{D.}~\bibnamefont{Avnir}},
  \bibinfo{journal}{Acta Cryst. B} \textbf{\bibinfo{volume}{60}},
  \bibinfo{pages}{163} (\bibinfo{year}{2004}), ISSN \bibinfo{issn}{0108-7681}.

\bibitem[{\citenamefont{Ok et~al.}(2006)\citenamefont{Ok, Halasyamani,
  Casanova, Llunell, Alemany, and Alvarez}}]{ok_distortions_2006}
\bibinfo{author}{\bibfnamefont{K.~M.} \bibnamefont{Ok}},
  \bibinfo{author}{\bibfnamefont{P.~S.} \bibnamefont{Halasyamani}},
  \bibinfo{author}{\bibfnamefont{D.}~\bibnamefont{Casanova}},
  \bibinfo{author}{\bibfnamefont{M.}~\bibnamefont{Llunell}},
  \bibinfo{author}{\bibfnamefont{P.}~\bibnamefont{Alemany}}, \bibnamefont{and}
  \bibinfo{author}{\bibfnamefont{S.}~\bibnamefont{Alvarez}},
  \bibinfo{journal}{Chem. Mater.} \textbf{\bibinfo{volume}{18}},
  \bibinfo{pages}{3176} (\bibinfo{year}{2006}).

\bibitem[{\citenamefont{Bernal and Fowler}(1933)}]{bernal_theory_1933}
\bibinfo{author}{\bibfnamefont{J.~D.} \bibnamefont{Bernal}} \bibnamefont{and}
  \bibinfo{author}{\bibfnamefont{R.~H.} \bibnamefont{Fowler}},
  \bibinfo{journal}{J. Chem. Phys.} \textbf{\bibinfo{volume}{1}},
  \bibinfo{pages}{515} (\bibinfo{year}{1933}).

\bibitem[{\citenamefont{Larson and Von~Dreele}(2000)}]{larson_general_2000}
\bibinfo{author}{\bibfnamefont{A.}~\bibnamefont{Larson}} \bibnamefont{and}
  \bibinfo{author}{\bibfnamefont{R.}~\bibnamefont{Von~Dreele}},
  \bibinfo{journal}{Los Alamos National Laboratory Report LAUR}
  \textbf{\bibinfo{volume}{86}}, \bibinfo{pages}{748} (\bibinfo{year}{2000}).

\bibitem[{\citenamefont{Peterson et~al.}(2000)\citenamefont{Peterson, Gutmann,
  Proffen, and Billinge}}]{peterson_pdfgetn_2000}
\bibinfo{author}{\bibfnamefont{P.~F.} \bibnamefont{Peterson}},
  \bibinfo{author}{\bibfnamefont{M.}~\bibnamefont{Gutmann}},
  \bibinfo{author}{\bibfnamefont{T.}~\bibnamefont{Proffen}}, \bibnamefont{and}
  \bibinfo{author}{\bibfnamefont{S.~J.~L.} \bibnamefont{Billinge}},
  \bibinfo{journal}{J. Appl. Cryst.} \textbf{\bibinfo{volume}{33}},
  \bibinfo{pages}{1192} (\bibinfo{year}{2000}).

\bibitem[{\citenamefont{Farrow et~al.}(2007)\citenamefont{Farrow, Juhas, Liu,
  Bryndin, Bozin, Bloch, Proffen, and Billinge}}]{farrow_pdffit2_2007}
\bibinfo{author}{\bibfnamefont{C.~L.} \bibnamefont{Farrow}},
  \bibinfo{author}{\bibfnamefont{P.}~\bibnamefont{Juhas}},
  \bibinfo{author}{\bibfnamefont{J.~W.} \bibnamefont{Liu}},
  \bibinfo{author}{\bibfnamefont{D.}~\bibnamefont{Bryndin}},
  \bibinfo{author}{\bibfnamefont{E.~S.} \bibnamefont{Bozin}},
  \bibinfo{author}{\bibfnamefont{J.}~\bibnamefont{Bloch}},
  \bibinfo{author}{\bibfnamefont{T.}~\bibnamefont{Proffen}}, \bibnamefont{and}
  \bibinfo{author}{\bibfnamefont{S.~J.~L.} \bibnamefont{Billinge}},
  \bibinfo{journal}{J. Phys. Cond. Mat.} \textbf{\bibinfo{volume}{19}},
  \bibinfo{pages}{335219} (\bibinfo{year}{2007}).

\bibitem[{\citenamefont{Kresse and Hafner}(1993)}]{kresse_ab_1993}
\bibinfo{author}{\bibfnamefont{G.}~\bibnamefont{Kresse}} \bibnamefont{and}
  \bibinfo{author}{\bibfnamefont{J.}~\bibnamefont{Hafner}},
  \bibinfo{journal}{Phys. Rev. B} \textbf{\bibinfo{volume}{47}},
  \bibinfo{pages}{558} (\bibinfo{year}{1993}).

\bibitem[{\citenamefont{Kresse and
  Furthm\"{u}ller}(1996)}]{kresse_efficient_1996}
\bibinfo{author}{\bibfnamefont{G.}~\bibnamefont{Kresse}} \bibnamefont{and}
  \bibinfo{author}{\bibfnamefont{J.}~\bibnamefont{Furthm\"{u}ller}},
  \bibinfo{journal}{Phys. Rev. B} \textbf{\bibinfo{volume}{54}},
  \bibinfo{pages}{11169} (\bibinfo{year}{1996}).

\bibitem[{\citenamefont{Bl\"{o}chl}(1994)}]{blochl_projector_1994}
\bibinfo{author}{\bibfnamefont{P.}~\bibnamefont{Bl\"{o}chl}},
  \bibinfo{journal}{Phys. Rev. B} \textbf{\bibinfo{volume}{50}},
  \bibinfo{pages}{17953} (\bibinfo{year}{1994}).

\bibitem[{\citenamefont{Kresse and Joubert}(1999)}]{kresse_ultrasoft_1999}
\bibinfo{author}{\bibfnamefont{G.}~\bibnamefont{Kresse}} \bibnamefont{and}
  \bibinfo{author}{\bibfnamefont{D.}~\bibnamefont{Joubert}},
  \bibinfo{journal}{Phys. Rev. B} \textbf{\bibinfo{volume}{59}},
  \bibinfo{pages}{1758} (\bibinfo{year}{1999}).

\bibitem[{bil()}]{bilbao}
\emph{\bibinfo{title}{Bilbao crystallographic server}},
  \bibinfo{note}{\texttt{http://www.cryst.ehu.es}}.

\bibitem[{\citenamefont{H.~T.~Stokes and Campbell}(2007)}]{isotropy}
\bibinfo{author}{\bibfnamefont{D.~M.~H.} \bibnamefont{H.~T.~Stokes}}
  \bibnamefont{and} \bibinfo{author}{\bibfnamefont{B.~J.}
  \bibnamefont{Campbell}}, \emph{\bibinfo{title}{Isotropy}}
  (\bibinfo{year}{2007}),
  \bibinfo{note}{\texttt{http://stokes.byu.edu/isotropy.html}}.

\bibitem[{\citenamefont{Izumi and Dilanian}(2002)}]{izumi_recent_2002}
\bibinfo{author}{\bibfnamefont{F.}~\bibnamefont{Izumi}} \bibnamefont{and}
  \bibinfo{author}{\bibfnamefont{A.}~\bibnamefont{Dilanian}},
  \emph{\bibinfo{title}{Recent Research Developments in Physics}},
  vol.~\bibinfo{volume}{3} (\bibinfo{publisher}{Transworld Research Network},
  \bibinfo{address}{Trivandrum, India}, \bibinfo{year}{2002}), ISBN
  \bibinfo{isbn}{81-7895-046-4}.

\bibitem[{\citenamefont{Tucker et~al.}(2007)\citenamefont{Tucker, Keen, Dove,
  Goodwin, and Hui}}]{tucker_rmcprofile_2007}
\bibinfo{author}{\bibfnamefont{M.~G.} \bibnamefont{Tucker}},
  \bibinfo{author}{\bibfnamefont{D.~A.} \bibnamefont{Keen}},
  \bibinfo{author}{\bibfnamefont{M.~T.} \bibnamefont{Dove}},
  \bibinfo{author}{\bibfnamefont{A.~L.} \bibnamefont{Goodwin}},
  \bibnamefont{and} \bibinfo{author}{\bibfnamefont{Q.}~\bibnamefont{Hui}},
  \bibinfo{journal}{J. Phys. Cond. Mat.} \textbf{\bibinfo{volume}{19}},
  \bibinfo{pages}{335218} (\bibinfo{year}{2007}).

\bibitem[{\citenamefont{Keen}(2001)}]{keen_comparison_2001}
\bibinfo{author}{\bibfnamefont{D.~A.} \bibnamefont{Keen}}, \bibinfo{journal}{J.
  Appl. Cryst.} \textbf{\bibinfo{volume}{34}}, \bibinfo{pages}{172}
  (\bibinfo{year}{2001}).

\bibitem[{\citenamefont{Momma and Izumi}(2008)}]{momma_vesta_2008}
\bibinfo{author}{\bibfnamefont{K.}~\bibnamefont{Momma}} \bibnamefont{and}
  \bibinfo{author}{\bibfnamefont{F.}~\bibnamefont{Izumi}}, \bibinfo{journal}{J.
  Appl. Cryst.} \textbf{\bibinfo{volume}{41}}, \bibinfo{pages}{653}
  (\bibinfo{year}{2008}).

\bibitem[{\citenamefont{Li}(2003)}]{li_atomeye_2003}
\bibinfo{author}{\bibfnamefont{J.}~\bibnamefont{Li}},
  \bibinfo{journal}{Modelling and Simul. Mater. Sci. Eng.}
  \textbf{\bibinfo{volume}{11}}, \bibinfo{pages}{173} (\bibinfo{year}{2003}).

\bibitem[{\citenamefont{Radosavljevic et~al.}(1998)\citenamefont{Radosavljevic,
  Evans, and Sleight}}]{radosavljevic_synthesis_1998}
\bibinfo{author}{\bibfnamefont{I.}~\bibnamefont{Radosavljevic}},
  \bibinfo{author}{\bibfnamefont{J.~S.~O.} \bibnamefont{Evans}},
  \bibnamefont{and} \bibinfo{author}{\bibfnamefont{A.~W.}
  \bibnamefont{Sleight}}, \bibinfo{journal}{J. Solid State Chem.}
  \textbf{\bibinfo{volume}{136}}, \bibinfo{pages}{63} (\bibinfo{year}{1998}).

\bibitem[{\citenamefont{Evans et~al.}(2003)\citenamefont{Evans, Howard, and
  Evans}}]{evans_alpha_2003}
\bibinfo{author}{\bibfnamefont{I.~R.} \bibnamefont{Evans}},
  \bibinfo{author}{\bibfnamefont{J.~A.~K.} \bibnamefont{Howard}},
  \bibnamefont{and} \bibinfo{author}{\bibfnamefont{J.~S.~O.}
  \bibnamefont{Evans}}, \bibinfo{journal}{J. Mater. Chem.}
  \textbf{\bibinfo{volume}{13}}, \bibinfo{pages}{2098} (\bibinfo{year}{2003}).

\bibitem[{\citenamefont{Cagnon et~al.}(2007)\citenamefont{Cagnon, Boesch,
  Finstrom, Nergiz, Keane, and Stemmer}}]{cagnon_microstructure_2007}
\bibinfo{author}{\bibfnamefont{J.}~\bibnamefont{Cagnon}},
  \bibinfo{author}{\bibfnamefont{D.~S.} \bibnamefont{Boesch}},
  \bibinfo{author}{\bibfnamefont{N.~H.} \bibnamefont{Finstrom}},
  \bibinfo{author}{\bibfnamefont{S.~Z.} \bibnamefont{Nergiz}},
  \bibinfo{author}{\bibfnamefont{S.~P.} \bibnamefont{Keane}}, \bibnamefont{and}
  \bibinfo{author}{\bibfnamefont{S.}~\bibnamefont{Stemmer}},
  \bibinfo{journal}{J. Appl. Phys.} \textbf{\bibinfo{volume}{102}},
  \bibinfo{pages}{044102} (\bibinfo{year}{2007}).

\bibitem[{\citenamefont{Jaynes}(1957{\natexlab{a}})}]{jaynes_information1_1957}
\bibinfo{author}{\bibfnamefont{E.~T.} \bibnamefont{Jaynes}},
  \bibinfo{journal}{Phys. Rev.} \textbf{\bibinfo{volume}{106}},
  \bibinfo{pages}{620} (\bibinfo{year}{1957}{\natexlab{a}}).

\bibitem[{\citenamefont{Jaynes}(1957{\natexlab{b}})}]{jaynes_information2_1957}
\bibinfo{author}{\bibfnamefont{E.~T.} \bibnamefont{Jaynes}},
  \bibinfo{journal}{Phys. Rev.} \textbf{\bibinfo{volume}{108}},
  \bibinfo{pages}{171} (\bibinfo{year}{1957}{\natexlab{b}}).

\bibitem[{\citenamefont{Sakata and Sato}(1990)}]{sakata_accurate_1990}
\bibinfo{author}{\bibfnamefont{M.}~\bibnamefont{Sakata}} \bibnamefont{and}
  \bibinfo{author}{\bibfnamefont{M.}~\bibnamefont{Sato}},
  \bibinfo{journal}{Acta Cryst. A} \textbf{\bibinfo{volume}{46}},
  \bibinfo{pages}{263} (\bibinfo{year}{1990}).

\bibitem[{\citenamefont{Sakata et~al.}(1990)\citenamefont{Sakata, Mori,
  Kumazawza, Takata, and Toraya}}]{sakata_electron_1990}
\bibinfo{author}{\bibfnamefont{M.}~\bibnamefont{Sakata}},
  \bibinfo{author}{\bibfnamefont{R.}~\bibnamefont{Mori}},
  \bibinfo{author}{\bibfnamefont{S.}~\bibnamefont{Kumazawza}},
  \bibinfo{author}{\bibfnamefont{M.}~\bibnamefont{Takata}}, \bibnamefont{and}
  \bibinfo{author}{\bibfnamefont{H.}~\bibnamefont{Toraya}},
  \bibinfo{journal}{J. Appl. Cryst.} \textbf{\bibinfo{volume}{23}},
  \bibinfo{pages}{526} (\bibinfo{year}{1990}).

\bibitem[{\citenamefont{Kumazawa et~al.}(1993)\citenamefont{Kumazawa, Kubota,
  Takata, Sakata, and Ishibashi}}]{kumazawa_meed_1993}
\bibinfo{author}{\bibfnamefont{S.}~\bibnamefont{Kumazawa}},
  \bibinfo{author}{\bibfnamefont{Y.}~\bibnamefont{Kubota}},
  \bibinfo{author}{\bibfnamefont{M.}~\bibnamefont{Takata}},
  \bibinfo{author}{\bibfnamefont{M.}~\bibnamefont{Sakata}}, \bibnamefont{and}
  \bibinfo{author}{\bibfnamefont{Y.}~\bibnamefont{Ishibashi}},
  \bibinfo{journal}{J. Appl. Cryst.} \textbf{\bibinfo{volume}{26}},
  \bibinfo{pages}{453} (\bibinfo{year}{1993}).

\bibitem[{\citenamefont{Avdeev et~al.}(2002)\citenamefont{Avdeev, Haas,
  Jorgensen, and Cava}}]{avdeev_static_2002}
\bibinfo{author}{\bibfnamefont{M.}~\bibnamefont{Avdeev}},
  \bibinfo{author}{\bibfnamefont{M.~K.} \bibnamefont{Haas}},
  \bibinfo{author}{\bibfnamefont{J.~D.} \bibnamefont{Jorgensen}},
  \bibnamefont{and} \bibinfo{author}{\bibfnamefont{R.~J.} \bibnamefont{Cava}},
  \bibinfo{journal}{J. Solid State Chem.} \textbf{\bibinfo{volume}{169}},
  \bibinfo{pages}{24} (\bibinfo{year}{2002}).

\bibitem[{\citenamefont{Tabira et~al.}(1999)\citenamefont{Tabira, Withers,
  Thompson, and Schmid}}]{tabira_structured_1999}
\bibinfo{author}{\bibfnamefont{Y.}~\bibnamefont{Tabira}},
  \bibinfo{author}{\bibfnamefont{R.}~\bibnamefont{Withers}},
  \bibinfo{author}{\bibfnamefont{J.}~\bibnamefont{Thompson}}, \bibnamefont{and}
  \bibinfo{author}{\bibfnamefont{S.}~\bibnamefont{Schmid}},
  \bibinfo{journal}{J. Solid State Chem.} \textbf{\bibinfo{volume}{142}},
  \bibinfo{pages}{393} (\bibinfo{year}{1999}).

\bibitem[{\citenamefont{Nguyen et~al.}(2007)\citenamefont{Nguyen, Liu, and
  Withers}}]{nguyen_local_2007}
\bibinfo{author}{\bibfnamefont{B.}~\bibnamefont{Nguyen}},
  \bibinfo{author}{\bibfnamefont{Y.}~\bibnamefont{Liu}}, \bibnamefont{and}
  \bibinfo{author}{\bibfnamefont{R.~L.} \bibnamefont{Withers}},
  \bibinfo{journal}{J. Solid State Chem.} \textbf{\bibinfo{volume}{180}},
  \bibinfo{pages}{549} (\bibinfo{year}{2007}).

\bibitem[{\citenamefont{Withers et~al.}(1989)\citenamefont{Withers, Thompson,
  and Welberry}}]{withers_structure_1989}
\bibinfo{author}{\bibfnamefont{R.}~\bibnamefont{Withers}},
  \bibinfo{author}{\bibfnamefont{J.}~\bibnamefont{Thompson}}, \bibnamefont{and}
  \bibinfo{author}{\bibfnamefont{T.}~\bibnamefont{Welberry}},
  \bibinfo{journal}{Phys. Chem. Miner.} \textbf{\bibinfo{volume}{16}},
  \bibinfo{pages}{517} (\bibinfo{year}{1989}).

\bibitem[{\citenamefont{Proffen and Welberry}(1997)}]{proffen_analysis_1997}
\bibinfo{author}{\bibfnamefont{T.}~\bibnamefont{Proffen}} \bibnamefont{and}
  \bibinfo{author}{\bibfnamefont{T.~R.} \bibnamefont{Welberry}},
  \bibinfo{journal}{Acta Cryst. A} \textbf{\bibinfo{volume}{53}},
  \bibinfo{pages}{202} (\bibinfo{year}{1997}).

\end{thebibliography}

\end{document}